\documentclass[a4paper,11pt]{article}
\pdfoutput=1
\usepackage{jcappub}
  
\usepackage{tabularx} 
\usepackage{arydshln}
\usepackage{amsmath,amssymb}  
\usepackage{physics}
\usepackage{bm}
\usepackage{graphicx}
\usepackage{enumitem}
\usepackage{geometry}
\usepackage{xspace}
\usepackage{pdflscape}
\usepackage{enumerate}
\makeatletter
\gdef\@fpheader{}
\g@addto@macro\bfseries{\boldmath}
\makeatother
\DeclareUnicodeCharacter{2212}{-}

\usepackage{soul}

\newcommand{\delr}{\boldsymbol{\nabla}_r}
\newcommand{\divr}{\boldsymbol{\nabla}_r\boldsymbol{\cdot}}
\newcommand{\calA}{{\cal B}}

\title{Observational constraints on interacting vacuum energy with linear interactions}

\author[a]{Chakkrit Kaeonikhom,}
\author[a,b]{Hooshyar Assadullahi,}
\author[a]{Jascha Schewtschenko,}
\author[a]{David Wands}

\affiliation[a]{Institute of Cosmology and Gravitation, University of Portsmouth,\\Dennis Sciama Building, Burnaby Road, Portsmouth, PO1 3FX, UK}

\affiliation[b]{School of Mathematics and Physics, University of Portsmouth,\\Lion Gate Building, Lion Terrace, Portsmouth, PO1 3HF, UK}

\emailAdd{chakkrit.kaeonikhom@port.ac.uk}
\emailAdd{hooshyar.assadullahi@port.ac.uk}
\emailAdd{jascha.schewtschenko@port.ac.uk}
\emailAdd{david.wands@port.ac.uk}

\abstract{We explore the bounds that can be placed on interactions between cold dark matter and vacuum energy, with equation of state $w=-1$, using state-of-the-art cosmological observations. We consider linear perturbations about a simple background model where the energy transfer per Hubble time, $Q/H$, is a general linear function of the dark matter density, $\rho_c$, and vacuum energy, $V$. 
We explain the parameter degeneracies found when fitting cosmic microwave background (CMB) anisotropies alone, and show how these are broken by the addition of supernovae data, baryon acoustic oscillations (BAO) and redshift-space distortions (RSD). In particular, care must be taken when relating redshift-space distortions to the growth of structure in the presence of non-zero energy transfer. Interactions in the dark sector can alleviate the tensions between low-redshift measurements of the Hubble parameter, $H_0$, or weak-lensing, $S_8$, and the values inferred from CMB data. However these tensions return when we include constraints from supernova and BAO-RSD datasets. In the general linear interaction model we show that, while it is possible to relax both the Hubble and weak-lensing tensions simultaneously, the reduction in these tensions is modest (reduced to less slightly than $4\sigma$ and $2\sigma$ respectively).}

\begin{document}

\maketitle
\flushbottom


\section{Introduction}

There is now strong evidence that the expansion of the Universe is accelerating from a variety of observations, since early measurements of the magnitude-redshift relation for type-Ia Supernovae (SNe Ia) \cite{Perlmutter:1998np,Riess:1998cb}, to detailed observations of cosmic microwave background (CMB) anisotropies made by Planck satellite \cite{Planck:2018vyg}, as well as galaxy surveys~\cite{Beutler:2012px,Howlett:2014opa,10.1093/mnras/stu111,Alam:2016hwk}. Such an acceleration suggests that the Universe contains some kind of \textit{dark energy}, homogeneously distributed exerting a negative pressure, in addition to both ordinary, baryonic matter and non-relativistic \textit{cold dark matter} (CDM) which freely falls under gravity. 
The simplest model for dark energy is a cosmological constant, $\Lambda$, with an effective equation of state $w=-1$, which is equivalent to a non-zero vacuum energy as predicted by quantum field theory. However, the huge discrepancy between the vacuum energy inferred from observations and that expected from theoretical predictions gives rise to the cosmological constant problem \cite{Weinberg:1989,Carroll:1991mt}.

The standard $\Lambda$CDM cosmology can successfully describe all current observations with suitable choices for the cosmological parameters. However the parameter values inferred from CMB anisotropies appear to be in tension with other determinations of the Hubble constant $H_0$ based distance-ladder measurements at lower redshifts. In particular, the SH0ES collaboration estimates $H_0=73.04\pm1.04 \;\mathrm{km\,s^{-1}Mpc^{-1}}$ based on observations of SNe calibrated by a sample of Cepheid variable stars \cite{Riess:2021jrx}. This gives rise to about $4.8\sigma$ statistically significant difference with respect to the Planck baseline CMB constraint, given by $H_0=67.36\pm0.54\;\mathrm{km\,s^{-1}Mpc^{-1}}$ \cite{Planck:2018vyg} in $\Lambda$CDM. 
If this discrepancy is not due to undiscovered systematic errors in either of the analyses, the only explanation for this Hubble tension appears to be new physics beyond the standard $\Lambda$CDM cosmology.
%
%
At the same time there is growing evidence of another tension between the amplitude of matter density fluctuations measured by weak-lensing and other surveys of large-scale structure, denoted by $\sigma_8$ or $S_8\equiv \sigma_8(\Omega_m/0.3)^{1/2}$, compared with the amplitude inferred from CMB anisotropies in $\Lambda$CDM. 
The most common approach to address these tensions by exploring cosmologies beyond $\Lambda$CDM is to consider a dynamical form of dark energy by introducing additional degrees of freedom in the form of a field or fluid description, leading to an effective equation of state $w\neq1$~\cite{Copeland:2006wr}. However most attempts to resolve the Hubble tension tend to exacerbate the discrepancy in $S_8$~\cite{Anchordoqui:2021gji}.

Here we investigate an alternative, less frequently studied extension of $\Lambda$CDM where the equation of state of dark energy remains fixed to be that of a vacuum energy, $w=-1$, but where we allow the vacuum to exchange energy-momentum with dark matter~\cite{Bertolami:1986bg,Freese:1986dd,Carvalho:1991ut,Shapiro:2000dz,Wands:2012vg,SolaPeracaula:2022hpd}. For example, one might consider vacuum decay due to particle creation in an expanding universe~\cite{Alcaniz:2012mh,Pigozzo:2015swa}.
Cosmological probes allow us to constrain the form of the interaction~\cite{Wang:2015wga,Hogg:2020rdp} while studying the effect of the interaction on other cosmological parameters such as $H_0$ and $S_8$.


%
%

%

In this paper we consider interacting vacuum cosmologies where the energy transfer per Hubble time, $Q/H=\alpha \rho_c+\beta V$, is a general linear function of the cold dark matter density, $\rho_c$, and/or vacuum energy, $V$, updating and extending previous studies~\cite{Sadjadi:2006qp,CalderaCabral:2008bx,Quartin:2008px,Quercellini:2008vh,Kaeonikhom:2020fqs}. This model generalises time-dependent vacuum models with $Q=\beta HV$~\cite{Salvatelli:2014zta,Martinelli:2019dau}, or time-dependent dark matter models with $Q=\alpha H\rho_c$~\cite{Shapiro:2003ui,EspanaBonet:2003vk,Wang:2004cp,Alcaniz:2005dg,Sola:2017jbl}, which are closely related to running vacuum models~\cite{SolaPeracaula:2022hpd}. It also generalises the decomposed Chaplygin gas models~\cite{Bento:2004uh,Wang:2013qy,Wang:2014xca,Wang:2018azy}, where $Q = \gamma H\rho_cV/(\rho_c+V)$, and we have $Q/H\to\gamma V$ at early times where $V/\rho_c\to0$, and $Q/H\to\gamma \rho_c$ at late times where $\rho_c/V\to0$.
%

This paper is organised as follows. In Section~\ref{sec:IVmodels} we introduce the general covariant equations for the vacuum-dark matter interaction and give analytic solutions for the background cosmological models for specific and general linear interaction models. In Section~\ref{sec:cpt} we present the dynamical equations for scalar cosmological perturbations, focusing on the geodesic model for inhomogeneous perturbations where dark matter follows geodesics and hence clusters while the vacuum energy remains homogeneous in the synchronous, comoving frame~\cite{Wands:2012vg}. The next Section~\ref{sec:cmb_aniso} investigates the effect of interaction parameters on the CMB power spectra and in Section~\ref{sec:rsd} we consider the interpretation of redshift-space distortions in this model. We present constraints on the model parameters from current observations of CMB anisotropies, type-Ia supernovae and large-scale structure survey data in Section~\ref{sec:observations} and discuss the impact on $H_0$ and $S_8$ tensions in Section~\ref{sec:discussion}. We present our conclusions in Section~\ref{conclusions}.



\section{Interacting vacuum energy models}
\label{sec:IVmodels}


\subsection{Covariant equations}

The energy-momentum tensor of the vacuum energy $V$ can be defined as proportional to the metric tensor
\begin{equation}
\label{eq:stress_tensor_V}
\check{T}_{\nu}^{\mu} = -V g_{\nu}^{\mu}
\end{equation}
with the energy density $\rho_{V}=V$ and pressure $P_{V}=-V$.
The equation of state parameter thus corresponds to that of a cosmological constant, $w=P_{V}/\rho_{V}=-1$.
Comparing Eq.~\eqref{eq:stress_tensor_V}
with the energy-momentum tensor for a perfect fluid,
\begin{equation}
T^{\mu}_{\nu} = Pg^{\mu}_{\nu} + (\rho+P)u^{\mu}u_{\nu} \,,
\end{equation}
we can see that the four-velocity for the vacuum, $\check{u}^{\mu}$, 
is not defined since $\rho_{V}+P_{V}=0$ and the 4-momentum
is identically zero~\cite{Wands:2012vg}.

In general relativity the total energy-momentum tensor of matter plus vacuum is covariantly conserved
$
\nabla_\mu\qty(T_\nu^{\mu} + \check{T}_\nu^{\mu} ) = 0 \,, 
$
however an energy-momentum exchange between these two components is allowed.
We define the energy-momentum transfer to the vacuum energy
\begin{equation}
Q_\nu \equiv 
\nabla_\mu \check{T}_\nu^{\mu}  = - \nabla_\nu V \,,
\end{equation}
where the final equality can be obtained directly from the definition of the vacuum energy-momentum tensor, Eq.~\eqref{eq:stress_tensor_V}.
Hence, from the conservation of the total energy-momentum, we have  
\begin{equation}
	\nabla_{\mu}T_{\nu}^{\mu} = -Q_\nu \,.
\end{equation}
Without loss of generality we can decompose the energy-momentum transfer~\cite{Kodama:1985bj,Malik:2004tf} into an energy transfer $Q$ along the matter 4-velocity, $u^\mu$, and a momentum transfer (or force), $f^\mu$ orthogonal to $u^{\mu}$:
\begin{equation}
    Q^\mu = Q u^\mu +f^\mu \,.
\end{equation}

%
%

\subsection{FLRW cosmology}

In a spatially-flat Friedmann-Lemaitre-Robertson-Walker (FLRW) background cosmology with vacuum energy $V$, the Friedmann equation reads
\begin{equation}
\label{eq:FriedmannEq}
	H^2 = \frac{8\pi G}{3}(\rho_r + \rho_b + \rho_c + V),
\end{equation} 
with the Hubble rate $H=\dot{a}/a$, where $a$ is the scale factor and a dot denotes a time derivative. The radiation and matter densities are denoted by $\rho_i$ where the subscripts $r$, $b$ and $c$ stand for radiation, baryons and cold dark matter respectively. 

In an FLRW background, all the matter components, including the vacuum energy, are spatially homogeneous. Hence, the energy transfer $Q$ only depends on time. In this case the FLRW symmetry required that the energy transfer four-vector is parallel to the dark matter four-velocity, $Q^\mu=Qu^\mu$.
%
The continuity equations take the simple form:
\begin{equation}
    \begin{aligned}
	\dot{V}&= Q\,,\\
	\dot{\rho}_{r} + 4H\rho_{r} &= 0\,,\\
	\dot{\rho}_{b} + 3H\rho_{b} &= 0\,,\\
	\dot{\rho}_{c} + 3H\rho_{c} &= -Q\,,
    \end{aligned}
\label{eq:idm_iv_fld_Q}
\end{equation}
 and we assume that the interaction between vacuum and matter is restricted to the dark sector. Radiation and baryons are only gravitationally coupled to the vacuum energy, giving the standard background solutions 
\begin{equation}
\label{rhobc}
    \rho_r = \rho_{r,0} a^{-4} \,, \quad \rho_b = \rho_{b,0} a^{-3} \,, 
\end{equation}

In this paper, we consider the simple linear interaction model~\cite{Sadjadi:2006qp,CalderaCabral:2008bx,Quartin:2008px,Quercellini:2008vh,Kaeonikhom:2020fqs}
\begin{equation}
	\label{eq:linear_model_Q}
	Q = \alpha H\rho_{c}+\beta HV\,,
\end{equation} 
where $\alpha$ and $\beta$ are dimensionless parameters controlling the strength of the interaction.
We will study separately the cases $\beta=0$ and $\alpha=0$, i.e., the models $Q=\alpha H\rho_c$ and $Q=\beta HV$, before considering the general case.


\subsection{Model: $Q=\alpha H\rho_{c}$}
\label{ssec:a_model}

In this model, Eqs.~\eqref{eq:idm_iv_fld_Q} can be integrated to give
\begin{align}
\rho_c &= \rho_{c,0}a^{-(3+\alpha)}, 
\label{Eq:bg_a_exact_rhocdm} \\
\rho_{V} &= \frac{\alpha}{\alpha +3}\left(1 - a^{-(3+\alpha)}\right)\rho_{c,0} + V_0,
\label{Eq:bg_a_exact_rhov}
\end{align}
where $\rho_{c,0}$ and $V_0$ are the present energy densities of cold dark matter and vacuum energy respectively.
The Friedmann equation, Eq.~\eqref{eq:FriedmannEq}, can then be rewritten in terms of the present-day dimensionless density parameters
\begin{equation}
    \Omega_{i} \equiv \frac{8\pi G\rho_{i}}{3H^2} \,,
\end{equation}
as
%
%
\begin{equation}
\label{Hzalpha}
	H^2(z) = H_{0}^{2}
	\left[\Omega_{r,0}(z+1)^{4} + \Omega_{b,0}(z+1)^{3} + \left\{\frac{\alpha + 3\,(z+1)^{\alpha +3}}{\alpha +3}\right\}\Omega_{c,0} + \Omega_{V,0}\right] \, .
\end{equation} 
where the redshift $z=a^{-1}-1$. Note that the total matter density is given by $\Omega_{m,0}=\Omega_{b,0}+\Omega_{c,0}$ and, by construction, $\Omega_{V,0}=1-\Omega_{r,0}-\Omega_{m,0}$. 

When exploring observational constraints on the expansion history, $H(z)$, and the dependence on different cosmological parameters it will be helpful to eliminate $\Omega_{V,0}$ and re-write Eq.~\eqref{Hzalpha} in terms of the parameters $\omega_b\equiv\Omega_b h^2$, $\omega_c\equiv\Omega_c h^2$ and $\omega_r\equiv\Omega_r h^2$, as well as $\alpha$ 
and the dimensionless Hubble constant, $h\equiv H_0/H_{100}$, where $H_{100}=100\,{\rm km\,s^{-1}\,Mpc^{-1}}$. 
This gives
\begin{align}
	H(z) &= H_{100}
	\bigg[{\omega}_{r}\left\{(1+z)^{4}-1\right\} + {\omega}_{b}\left\{(1+z)^{3}-1\right\} 
	+ \frac{3{\omega}_{c}}{3+\alpha}\left\{(1+z)^{3+\alpha}-1\right\} + h^2\bigg]^{1/2} \,.
	\label{eq:Hz_alpha}
\end{align}
%
%
For $\alpha=0$ we recover the standard expression for the $\Lambda$CDM model with the same parameter values
\begin{equation}
\label{barHz}
	{H}_\Lambda(z) = 
	H_{100} 
	\left[{\omega}_r\left\{(1+z)^4-1\right\}+ ({\omega}_b+{\omega}_c)\left\{(1+z)^3-1\right\} + h^2\right]^{1/2} \,.
\end{equation}

\subsection{Model: $Q=\beta HV$}

\label{ssec:b_model}

The solution of Eq.~\eqref{eq:idm_iv_fld_Q} for this interaction model can be written as
\begin{align}
\label{rhocbeta}
	\rho_c &= \rho_{c,0}(1+z)^3 + \calA(\beta,z)V_0\\
	V &= V_0 (1+z)^{-\beta}
\end{align}
where
\begin{equation}
\label{eq:A_function}
{\calA}(\beta,z) \equiv \frac{\beta}{\beta+3}\left\{(1+z)^3-(1+z)^{-\beta }\right\}
\end{equation}

The Friedmann equation, Eq.~\eqref{eq:FriedmannEq}, then can be written as 
%
\begin{align}
\label{Hzbeta}
 H^2(z) = H_0^2
\left[\Omega_{r,0}(1+z)^{4} + \Omega_{b,0}(1+z)^{3} + \Omega_{c,0}(1+z)^3 + 
\left\{{\calA}(\beta,z)+(1+z)^{-\beta}\right\}\Omega_{V,0}\right]
\,.
\end{align}
Eliminating $\Omega_{V,0}$ and writing the expansion history, $H(z)$, in terms of the physical density parameters, $\omega_b\equiv\Omega_b h^2$, $\omega_c\equiv\Omega_c h^2$ and $\omega_r\equiv\Omega_r h^2$, we can write Eq.~\eqref{Hzbeta} for the expansion history in the $\beta$-model as
\begin{align}
\label{Hzbetaomegas}
    H^2(z) &= 
    H_{100}^2 \bigg\{ \omega_r(1+z)^4 + (\omega_b+\omega_c)(1+z)^3 
	 + (h^2-\omega_r-\omega_b-\omega_c) \left[ (1+z)^{-\beta} + \calA(\beta,z)  \right] \bigg\} \,.
\end{align}
%
For $\beta=0$ we recover the usual $\Lambda$CDM expression \eqref{barHz}.

\subsection{Model: $Q=\alpha H\rho_{c}+\beta HV$}
\label{ssec:ab_model}

The analytic solutions for dark matter density and vacuum energy for the general linear interaction (with both $\alpha\neq0$ and $\beta\neq0$) are
%
%
%
%
%
%
%
%
\begin{align}
\label{rhocalphabeta}
    \rho_c &= C_+(1+z)^{p_+} - C_-(1+z)^{p_-} \,, \\ 
    V &= V_+(1+z)^{p_-} - V_-(1+z)^{p_+} \,, 
\end{align}
where we define
\begin{align}
    C_\pm &\equiv \frac{3+\alpha+\beta\pm S}{2S}\rho_{c,0} + \frac{\beta}{S}V_0 \,, \\
    V_\pm &\equiv \frac{\alpha}{S}\rho_{c,0} + \frac{3+\alpha+\beta\pm S}{2S}V_0  \,, \\
    p_\pm &\equiv \frac{3+\alpha-\beta\pm S}{2} \,, \\
    S &\equiv \sqrt{(\alpha+\beta + 3)^2 - 4\alpha\beta} \,.
\end{align}
The Friedmann equation, Eq.~\eqref{eq:FriedmannEq}, becomes 
%
\begin{align}
\label{Hzalphabeta}
 H^2(z) &= H_0^2
\bigg[\Omega_{r,0}(1+z)^{4} + \Omega_{b,0}(1+z)^{3} 
\nonumber \\
&
\quad \quad \left. + 
\Omega_{c,0} \left\{ 
\frac{3-\alpha+\beta+S}{2S} (1+z)^{p_+} 
- \frac{3-\alpha+\beta-S}{2S} (1+z)^{p_-} 
\right\} \right.
\nonumber \\
&
\quad \quad
+ 
\Omega_{V,0} \left\{ 
\frac{3+\alpha-\beta+S}{2S} (1+z)^{p_-} 
- \frac{3+\alpha-\beta-S}{2S} (1+z)^{p_+} 
\right\} 
\bigg]
\,.
\end{align}

\section{Linear perturbations}
\label{sec:cpt}

We consider linear perturbations of a spatially-flat FLRW universe in an arbitrary gauge \cite{Kodama:1985bj,Mukhanov:1990me,Malik:2004tf,Malik:2008im}
\begin{equation}
    \dd{s}^2 = -(1+2A)\dd{t}^2 + 2a\partial_iB\dd{t}\dd{x}^i + a^2\qty[(1+2C)\delta_{ij}+2\partial_i\partial_jE]\dd{x}^i\dd{x}^j.
\end{equation}
where $A$, $B$, $C$ and $E$ are scalar metric perturbations.
The total four-velocity of matter (neglecting vorticity) is given by
\begin{equation}
\label{u}
 u^\mu = \left[1-A \,, a^{-1}\partial^i v \right] \,, \quad u_\mu = \left[ -1-A \,, \partial_i \theta \right] \,.
\end{equation}
where we define 
$\theta=a(v+B)$.

%


It will be helpful to identify a few gauge-invariant quantities that are commonly used~\cite{Malik:2008im}:
\begin{itemize}
\item{Comoving curvature perturbation:
\begin{equation}
\label{def:R}
{\cal R} = C + H\theta \,,
\end{equation}
}
\item{Bardeen/Newtonian potential:
\begin{equation}
\label{def:Psi}
\Psi = - C + H\sigma \,,
\end{equation}
}
\item{Eulerian-frame velocity:
\begin{equation}
\label{def:vartheta}
\vartheta = \theta + \sigma
\end{equation}
}
\end{itemize}
where the shear of the constant-time hypersurfaces is given by $\sigma_{ij}\equiv (\partial_i\partial_j-(1/3)\delta_{ij}\nabla^2) \sigma$ and $\sigma\equiv a^2\dot{E}-aB$. Note that these gauge-invariant quantities are not independent and we have
\begin{equation}
\label{RfromPsi}
{\cal R} = - \Psi + H\vartheta \,.
\end{equation}

\subsection{Energy continuity and Euler equations}

The continuity equations for first-order perturbations of the baryons, dark matter and vacuum energy read~\cite{Malik:2008im,Wands:2012vg}
\begin{align}
\label{evol:deltarhob}
 \dot{\delta\rho}_b + 3H\delta\rho_b
 +3 \rho_b \dot{C} + \rho_b\frac{\nabla^2}{a^2} \left( \theta_b + \sigma \right)
 &= 0
 \,,\\
\label{evol:deltarhoc}
 \dot{\delta\rho}_c + 3H\delta\rho_c
 +3 \rho_c \dot{C} + \rho_c\frac{\nabla^2}{a^2} \left( \theta_c + \sigma \right)
 &= -\delta Q - QA
 \,,\\
 \label{evol:deltaV}
 \dot{\delta V} &= \delta Q + QA \,,
\end{align}
while momentum conservation requires 
\begin{align}
\label{evol:thetab}
\rho_b \left( \dot\theta_b  + A \right) &= 0
 \,,\\
\label{evol:thetac}
\rho_c \left( \dot\theta_c  +A \right) -Q\theta_c  &= - f - Q\theta
 \,,\\
\label{evol:thetaV}
 -\delta V &= f + Q\theta \,,
\end{align}
where following \cite{Kodama:1985bj,Malik:2004tf}
we decompose $Q^\mu  = Q u^\mu + f^\mu$, where $f_\mu u^\mu=0$, so that
\begin{equation}
 Q_\mu = \left[ -Q(1+A)-\delta Q \,, \partial_i (f + Q\theta) \right] \,,
 \end{equation}
and we define the total matter momentum $(\rho_b+\rho_c)\theta\equiv \rho_b\theta_b+\rho_c\theta_c$.

The baryon and dark matter equations are commonly written in terms of the dimensionless density contrasts
\begin{equation}
\label{def:delta}
\delta_b = \frac{\delta\rho_b}{\rho_b},\quad \delta_c = \frac{\delta\rho_c}{\rho_c},
\end{equation}
the equations for baryons and dark matter can be written as
\begin{align}
\label{evol:deltab}
 \dot{\delta}_b  + 3 \dot{C} + \frac{\nabla^2}{a^2} \left( \theta_b + \sigma \right) &= 0
 \,,\\
\label{evol:deltac}
 \dot{\delta}_c  +3 \dot{C} + \frac{\nabla^2}{a^2} \left( \theta_c + \sigma \right)
 &= - \left( \frac{ \delta Q - Q(\delta_c-A)}{\rho_c} \right)\,.
\end{align}

Because the momentum of the vacuum is identically zero,
the conservation equation for the vacuum momentum, Eq.~\eqref{evol:thetaV}, becomes a constraint equation which requires that the vacuum pressure gradient, $\nabla_i(-\delta V)$ must be balanced by the force $\nabla_i(f+Q\theta)$. This in turn determines the equal and opposite force exerted by the vacuum on the dark matter
\begin{equation}
- \nabla_i \left( f +Q \theta \right) = \nabla_i \delta V \,.
 \end{equation}
Thus the dark matter particles feel a force due to the gradient of the vacuum potential energy.
We can use the vacuum energy and momentum conservation Eqs.~\eqref{evol:deltaV} and~(\ref{evol:thetaV}) to eliminate $\delta Q$ and $f$ in the dark matter Eqs.~\eqref{evol:deltarhoc} and~\eqref{evol:thetac} to obtain
\begin{align}
 \label{finaldeltarho}
 \dot{\delta\rho}_c + 3H\delta\rho_c
 +3 \rho_c\dot{C} +\rho_c \frac{\nabla^2}{a^2} \left( \theta + \sigma \right)
 = - \dot{\delta V}
 \,,\\
 \label{finaltheta}
\rho_c \left( \dot\theta_c + A \right) = \delta V + \dot{V} \theta_c
 \,.
\end{align}
In terms of the density contrast Eq.~\eqref{def:delta} we have~\cite{Salzano:2021zxk}
\begin{equation}
\dot\delta_c +3\dot{C} + \frac{\nabla^2}{a^2} \left( \theta_c + \sigma \right) = - \left( \frac{ \dot{\delta V} - \dot{V}\delta_c}{\rho_c} \right)\,.
\end{equation}

\subsection{Einstein equations}

The curvature and shear perturbations obey the evolution equations, driven by the isotropic pressure perturbation~\cite{Malik:2008im,Wands:2012vg}
\begin{align}
\ddot{C}+3HC-H\dot{A} - \left( 3H^2+2\dot{H}\right) A &= - 4\pi G \delta V \,,\\
\dot\sigma + H\sigma - A - C &= 0\,,
\end{align}
while the energy and momentum perturbations obey the Einstein constraint equations
\begin{align}
 \label{energy-con}
3H\left( \dot{C} - HA \right) - \frac{\nabla^2}{a^2} \left[ C - H \sigma \right] &= 4\pi G \left( \delta\rho_b +  \delta\rho_c + \delta V \right)
 \,,
\\
 \label{mtm-con}
\dot{C} - HA &= 4\pi G \left( \rho_b \theta_b + \rho_c\theta_c \right)
 \,.
 \end{align}
The vacuum contributes to the energy constraint Eq.~\eqref{energy-con} but not the momentum constraint Eq.~\eqref{mtm-con}.

\subsection{Geodesic model}

If the 4-vector $Q^\mu$ describing the energy-momentum transfer between the vacuum and the dark matter is parallel to the 4-velocity of the dark matter, $u^\mu$, then the 3-force, $\nabla_i f$, vanishes in the frame comoving with the dark matter:
\begin{equation}
f_c \equiv f + Q( \theta - \theta_c ) = 0 \,.
\end{equation}
We refer to this as a geodesic model, since the dark matter then follows geodesic.

The 3-force on the dark matter always vanishes in the FRW background due to isotropy. Hence the first-order perturbation, $\nabla_i f$, is not determined by the background solution for the energy transfer, $Q$. Rather, we need a covariant description of the inhomogeneous energy-momentum transfer, $Q^\mu$. 
If, for example, we require that the matter+vacuum perturbations are adiabatic in the sense that the relative (entropy) perturbation vanishes ($\delta V/\dot{V}=\delta\rho_c/\dot\rho_c$) then the interacting vacuum+matter behaves like a barotropic fluid, whose speed of sound is given by the adiabatic sound speed, $c_s^2=\dot{P}/\dot\rho$. This is non-zero (possibly imaginary), leading to oscillations (or instabilities) in the matter which place very strong constraints on the allowed interaction strength (see for example \cite{Wang:2013qy}).
By contrast, if we assume the 3-force, $\nabla_i f$, vanishes in the frame comoving with the dark matter, then the sound speed also vanishes, so that dark matter clusters on all scales, similar to non-interacting cold dark matter. Thus in this paper we will study the geodesic model as the minimal generalisation of a non-interacting $\Lambda$CDM cosmology.

In the geodesic case case
the momentum conservation Eqs.~(\ref{evol:thetab}--\ref{evol:thetaV}) simplify considerably to give
\begin{align}
\label{geoevol:thetab}
\dot\theta_b  + A  &= 0
 \,,\\
\label{geoevol:thetac}
\dot\theta_c  +A  &= 0 \,,\\
\label{geoevol:thetaV}
 -\delta V &= \dot{V} \theta_c \,.
\end{align}
In particular we find that the velocity potential for the baryons relative to the dark matter is constant,
\begin{equation}
\Theta_{b|c} \equiv \theta_b - \theta_c = {\rm constant} \,.
\end{equation}



We can exploit the spatial gauge freedom to choose a frame comoving with the dark matter ($\theta_c=0$) and choose a temporal gauge such that the spatial hypersurfaces are orthogonal to the dark matter four-velocity ($B=0$). From Eq.~(\ref{geoevol:thetac}), this gauge is then synchronous ($A=0$) and, from Eq.~(\ref{geoevol:thetaV}), the vacuum energy is spatially homogeneous ($\delta V=0$).
From Eq.~(\ref{evol:deltaV}) we then see that $\delta Q=0$ in this frame.

The energy continuity Eqs.~\eqref{evol:deltab} and~\eqref{finaldeltarho} are then
\begin{align}
\label{synevol:deltab}
\dot{\delta}_b &= -\frac{\dot{h}}{2}
 \,,\\
\label{synevol:deltac}
\dot\delta_c &= -\frac{\dot{h}}{2}+\frac{Q}{\rho_c}\delta_c \,,
\end{align}
where the perturbation of the trace of the metric in synchronous gauge is denoted by
\begin{equation}
h \equiv 6C+2\nabla^2 E \,,
\end{equation}
and the energy and momentum constraints, Eqs.~(\ref{energy-con}) and~(\ref{mtm-con}), reduce to 
\begin{align}
 \label{synenergy-con}
3H \dot{C} - \frac{\nabla^2}{a^2} \left[ C - H \sigma \right] &= 4\pi G \left( \rho_b\delta_b +  \rho_c\delta_c \right)
 \,,
\\
 \label{synmtm-con}
\dot{C} &= 4\pi G \rho_b \Theta_{b|c}
 \,.
\end{align}
%
%
%
If the baryons are initially comoving with the dark matter then $\Theta_{b|c}=0$ and the baryons follow the dark matter at all times in the frame comoving with the dark matter.
%
%
In this case, from Eq.~\eqref{synmtm-con}, the comoving curvature perturbation, Eq.~\eqref{def:R}, is a constant:
\begin{equation}
C = {\cal R} = {\rm constant} \,.
\end{equation} 

Finally, note that the shear in the comoving frame can be identified with the velocity in the (zero-shear) Eulerian frame, Eq.~\ref{def:vartheta}, $\vartheta=\sigma$.
The remaining equations of motion reduce to a set of coupled first-order differential equations for the density contrasts and the Eulerian velocity, 
\begin{align}
\label{comsynevol:deltab}
 \dot{\delta}_b  + \frac{\nabla^2}{a^2} \vartheta &= 0
 \,,\\
 \label{comsynevol:deltac}
\dot\delta_c + \frac{\nabla^2}{a^2} \vartheta &= \frac{Q}{\rho_c}\delta_c 
\,,\\
\label{synevol:sigma}
\dot\vartheta  &= - \Psi 
\,.
\end{align}
These are the Newtonian equations for the linear growth of structure that include the effect of the energy interaction between the dark matter and the vacuum, assuming a geodesic model, $f=0$, and assuming that baryons are initially comoving with the dark matter, $\Theta_{b|c}=0$.
In terms of the total matter density contrast we have
\begin{equation}
\label{comsynevol:delta}
\dot\delta + \frac{\nabla^2}{a^2} \vartheta = \left( \frac{Q}{\rho_b+\rho_c} \right) \delta \,.
\end{equation} 
These evolution equations are subject to the comoving energy constraint equation
\begin{align}
\label{comsynPoisson}
\frac{\nabla^2}{a^2} \Psi
&= 4\pi G \rho_m\delta
\,,
\end{align}
which we identify with the relativistic Poisson equation for the Newtonian potential.

If we differentiate the continuity equations for the density contrasts Eqs. \eqref{comsynevol:deltab}, \eqref{comsynevol:deltac} and  \eqref{comsynevol:delta}, and substitute in for the shear and its derivative using Eqs.~\eqref{synevol:sigma} and \eqref{comsynPoisson}, then we obtain second-order differential equations for each of the density contrasts~\cite{Salzano:2021zxk}:
\begin{align}
\ddot\delta_b + 2H \dot\delta_b &= 4\pi G \rho_m \delta \,,\\
\ddot\delta_c + \left( 2H - \frac{Q}{\rho_c} \right) \dot\delta_c - \left( \frac{\rho_c(\dot{Q}+5HQ)+Q^2}{\rho_c^2} \right) \delta_c &= 4\pi G \rho_m \delta \,,
\label{eq:growth_delta_cdm}\\
\ddot\delta + \left( 2H - \frac{Q}{\rho_m} \right) \dot\delta - \left( \frac{\rho_m(\dot{Q}+5HQ)+Q^2}{\rho_m^2} + 4\pi G \rho_m  \right) \delta &= 0 \,.
\label{eq:growth_delta_m}
\end{align}
%


\section{Cosmic microwave background anisotropies}
\label{sec:cmb_aniso}

We have used the background FLRW solutions and relativistic perturbation equations described in the preceding sections to include the interaction between dark matter and the vacuum in a modified version of the Boltzmann code \texttt{CLASS}. This enables us to investigate numerically the effect of the interaction on anisotropies in the cosmic microwave background radiation. We can identify two principal ways in which the strength of the vacuum-matter interaction is constrained by the CMB.

The CMB angular power spectrum fixes precisely the angular size of the sound horizon at recombination on the surface of last scattering, $\theta_{*}$, \cite{Planck:2018vyg}
\begin{equation}
\label{def:theta*}
100\theta_{*} = 1.04109 \pm 0.00030 
\,.
\end{equation}
It is determined by the ratio of the sound horizon scale at recombination, $r_{*}$, to the angular diameter distance, $D_A$, \cite{Hu:2001bc}
\begin{equation}
\label{def:theta_s}
\theta_{*}
= 
\frac{r_{*}}{D_{A*}}
\,.
\end{equation}
The sound horizon at recombination is given by
\begin{equation}
\label{eq:rs_rec}
r_{*} = 
 \int_{z_{*}}^{\infty}\frac{c_s(z)}{H(z)}\dd{z} \,,
\end{equation}
where 
$c_s(z)$ is the sound speed of the baryon-photon plasma which depends on the baryon-to-photon ratio determined by
\begin{equation}
    c_s(z) = \frac{c}{\sqrt{3(1+R(z))}},
\end{equation}
where $R(z)\equiv(3/4)\rho_b(z)/\rho_\gamma(z)$, and the angular diameter distance to the surface of last-scattering is given by
\begin{equation}
\label{def:dA_to_rec}
D_{A*}
= 
\int_{0}^{z_{*}}\frac{c \dd{z}}{H(z)}.
\end{equation}
Note that the redshift of recombination, $z_{*}\simeq 1090$, depends only very weakly on the nature of dark energy and thus on the interaction parameters, $\alpha$ and $\beta$; thus $z_{*}$ can be considered fixed. However the expansion history both before and after recombination does depend on the interaction between dark matter and the vacuum energy.  

There is a potential degeneracy between the sound horizon size, $r_{*}$, and the angular diameter distance, $D_{A*}$, that leaves $\theta_{*}$ in Eq.~\eqref{def:theta_s} invariant. 
However the height of the first acoustic peak in the CMB power spectrum is sensitive to the density of dark matter at the time of recombination, due to the effect of radiation driving~\cite{Hu:2001bc}. During the radiation era the amplitude of the acoustic oscillations is enhanced, whereas once matter dominates the gravitational potential this ends. Hence the height of the acoustic peaks in the CMB angular power spectrum constrains the density of dark matter at the time of recombination. If we know the matter density at recombination, and the radiation density at $z_{*}$ is fixed by the CMB temperature observed today, then we find that the sound horizon, $r_{*}$, is effectively fixed. 

The only remaining degeneracy in the angular acoustic scale, $\theta_{*}$, then comes from the degeneracy between the Hubble scale today, $H_0=100h\,{\rm km\,s^{-1}\,Mpc^{-1}}$, and the expansion history, $H(z)$, that leaves $D_{A*}$ in Eq.~\eqref{def:dA_to_rec} fixed. In particular $H(z)$ depends on both the matter-vacuum interaction parameters and the present matter density $\omega_c \equiv \Omega_{c,0} h^2$.

In $\Lambda$CDM the dark matter density at recombination, $\rho_{c*}$, is fixed once we specify the present day physical density $\omega_c \equiv \Omega_{c,0}h^2$, assuming we know the redshift of recombination, $z_{*}$. Thus we can use the height of the first acoustic peak to fix the dark matter density at recombination
%
\begin{equation}
\label{omegacLCDM}
\omega_{c*}
=\bar{\omega}_{c}(1+z_*)^3,
\end{equation}
where $\omega_{c*}\equiv 8\pi G\rho_{c*}/3H_{100}^2$
and we define the Planck 2018 best-fit value for $\omega_c$ in $\Lambda$CDM to be $\bar{\omega}_{c}\equiv\Omega_{c,0}h^2|_{\Lambda\rm CDM}$. In a model where the dark matter exchanges energy with the vacuum then the relationship between the present matter density and that at recombination is altered, leading to a degeneracy between the present matter density, $\omega_c$, and the interaction strength while leaving the dark matter density at recombination fixed. 

We discuss the combined effect of measurements the angular scale and height of the first acoustic peak for each of our interacting vacuum models in the following subsections. In practice we will use the full angular power spectra for TT,TE,EE+lowE+lensing obtained using the modified Boltzmann code \texttt{CLASS} in our full parameter constraints later in this paper.

\subsection{Model: $Q=\alpha H\rho_c$}

\begin{figure}
\centering
\includegraphics[width=1\textwidth]{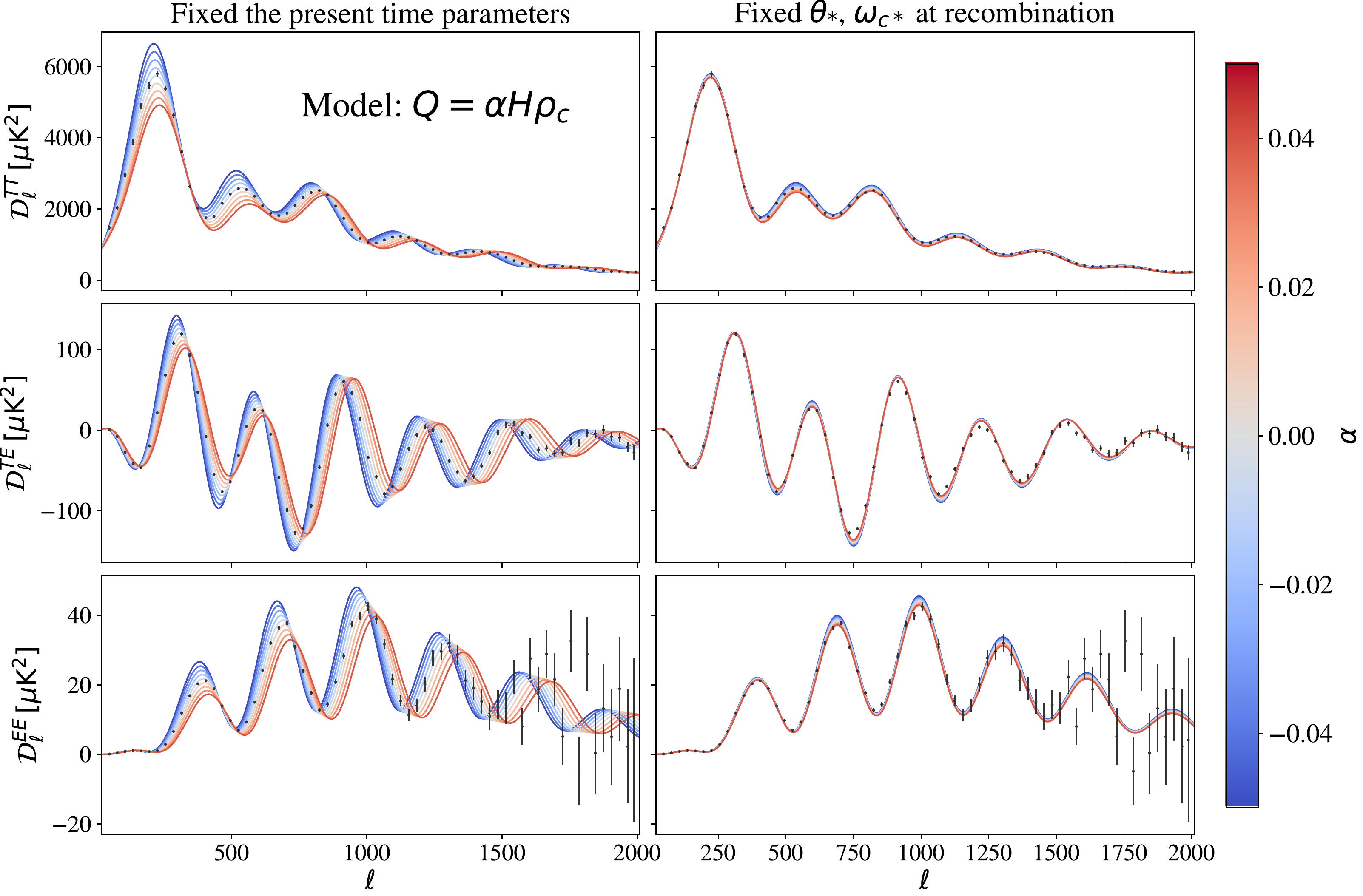}
\caption{\label{fig:Cls_alpha}The left panel shows the effect on the CMB power spectra in an interacting vacuum model of varying $\alpha$ while holding the other cosmological parameters ($\omega_b$, $\omega_c$ and $h$) fixed. The right panel shows the CMB power spectra for varying $\alpha$ while fixing the CMB angular acoustic scale, $\theta_*$, and the physical dark matter density parameter at recombination, $\omega_{c*}$, but allowing $h$ and $\omega_{c}$ to vary.}
\end{figure}

The physical dark matter density at recombination, $\omega_{c*}\equiv \Omega_{c*}h^2$ in the $\alpha$-model is related to the present-day density, $\omega_c$, for a given $\alpha$ by Eq.~\eqref{Eq:bg_a_exact_rhocdm}, giving
\begin{equation}
\label{omegac*}
\omega_{c*}
= \omega_{c}(1+z_*)^{\alpha+3} \,.
\end{equation}
%
Thus for a fixed value of $\omega_{c*}$ the matter density at recombination is larger for positive $\alpha$ and radiation driving is weaker. The amplitude of the acoustic peaks will therefore be smaller for positive $\alpha$, as we can see in the CMB power spectra on the left panel of Figure~\ref{fig:Cls_alpha}, where these plots use Planck 2018 best-fit values for the other cosmological parameters in $\Lambda$CDM based on $Planck$ TT,TE,EE+lowE+lensing shown in Table~\ref{tab:P18_six_param}.

%


\begin{table}
\centering
\caption{\label{tab:P18_six_param}\textit{Planck} 2018 baseline results taken from the base-$\Lambda$CDM model in the Table 2 of \cite{Planck:2018vyg}, which combines Planck CMB power spectra TT,TE,EE+lowE+lensing.}
\begin{tabular}{lc}
\hline\hline
Parameter                                       & TT,TE,EE+lowE+lensing \\
\hline
$\omega_b \equiv \Omega_{b} h^2$\dotfill                        & $0.02237\pm0.00015$ \\
$\omega_c \equiv \Omega_{c} h^2$\dotfill                        & $0.1200\pm0.0012$  \\
$\tau$\dotfill                                  & $0.0544\pm0.0073$   \\
$\ln(10^{10} A_{s})$\dotfill                    & $3.044\pm0.014$     \\
$n_{s}$\dotfill                                 & $0.9649\pm0.0042$   \\
$H_0\,[{\rm km}\,{\rm s}^{-1}\,{\rm Mpc}^{-1}]$ & $67.32\pm0.54$      \\
\hline
\end{tabular}

\end{table}

In practice the observed height of the acoustic peaks requires the dark matter density at recombination to remain close to the same value inferred in a $\Lambda$CDM cosmology, Eq.~\eqref{omegacLCDM}. Thus we have a relation between the physical dark matter density today in an interacting vacuum model and the best-fit value, $\bar{\omega}_c$, in $\Lambda$CDM
\begin{equation}
\label{fixomegacalpha}
\omega_c = \bar{\omega}_c(1+z_*)^{-\alpha}\,.
\end{equation}
For sufficiently small $\alpha$ we can expand this as
\begin{equation}
\label{degen:omegac-alpha}
\omega_c \simeq \bar{\omega}_c(1-7.0\alpha) \,,
\end{equation}
where we have taken $\ln(1+z_*)\simeq7.0$ at recombination.

\begin{figure}
\centering
\includegraphics[width=0.7\textwidth]{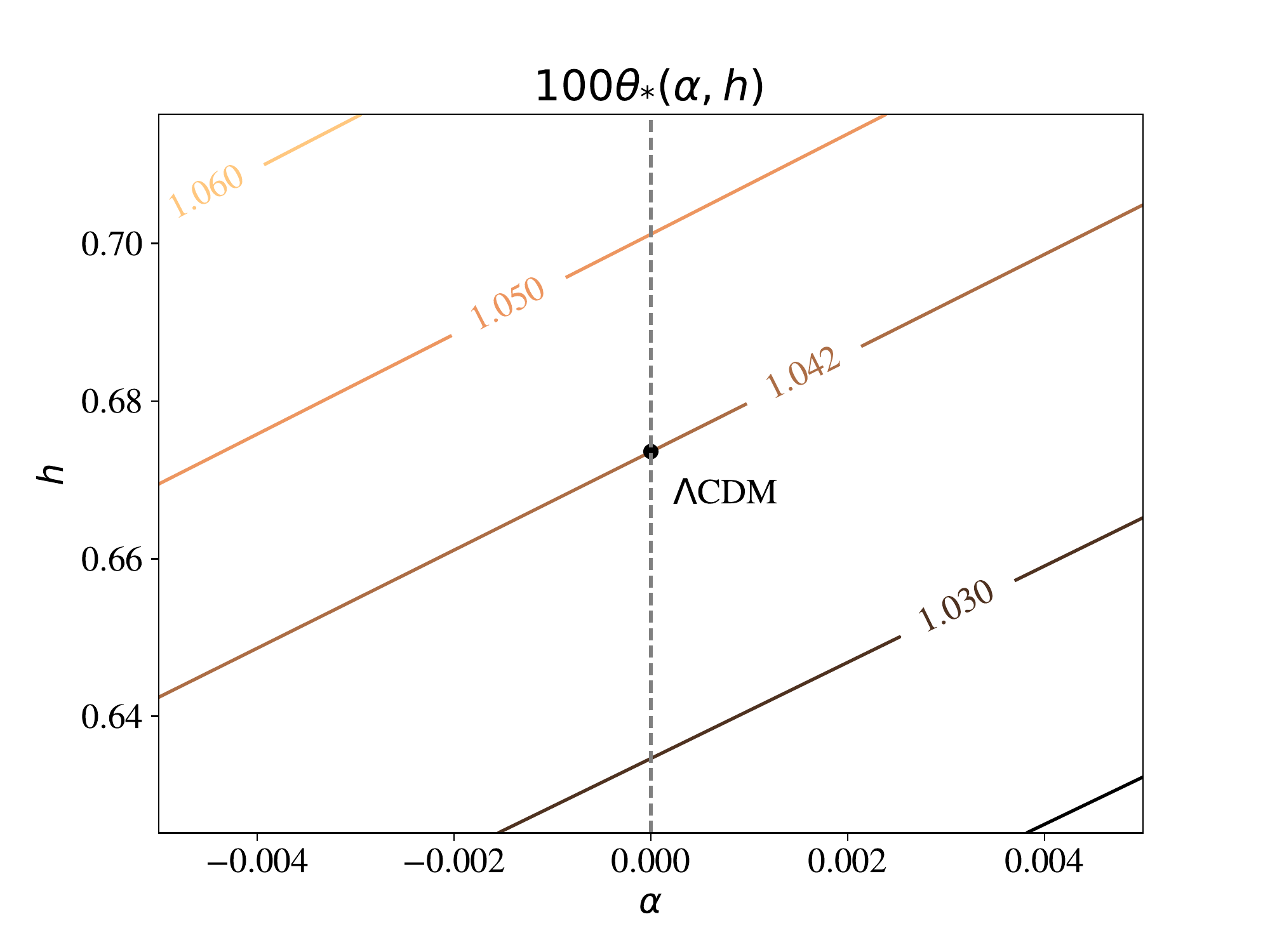}
\caption{\label{fig:alpha-h_theta} Contour plot showing the angular acoustic scale, $\theta_{*}$, for different values of $h$ and $\alpha$ while fixing the dark matter density at the time of recombination according to Eq.~\eqref{fixomegacalpha}. The middle diagonal line corresponds to the best-fit value of $\theta_*$ in the $\Lambda$CDM model.}
\end{figure}

The full dependence of the expansion history $H(z)$ on the cosmological parameters $\omega_b\equiv\Omega_b h^2$, $\omega_c\equiv\Omega_c h^2$ and $\omega_r\equiv\Omega_r h^2$, as well as $\alpha$ 
and the dimensionless Hubble constant, $h$, is given in Eq.~(\ref{eq:Hz_alpha}).
For sufficiently small $\alpha$, this
can be approximated by
\begin{equation}
\label{Hzsmallalpha1} 
	H^{2}(z)\simeq
	{H}_\Lambda^{2}(z)+
	H_{100}^2\,
	{\omega}_{c}
\left[ \left( \ln(1+z)-\frac{1}{3} \right) (1+z)^{3} + \frac{1}{3} \right] \alpha 
\,,
\end{equation}
where $H_\Lambda(z)$ corresponds to the $\Lambda$CDM expression given in Eq.~\eqref{barHz}.
We find that for small $\alpha$ there is a degeneracy in $H(z)$ between $\alpha$ and $h$, where the other parameters ($\omega_b$, $\omega_c$ and $\omega_r$) are held fixed in Eqs.~\eqref{Hzsmallalpha1} and~\eqref{barHz} for $H_\Lambda(z)$.
%
If we now fix the dark matter density at recombination, $\omega_{c*}$ in Eq.~\eqref{omegac*}, to coincide with that in the best-fit $\Lambda$CDM model then the constrained expansion history can be written, for small $\alpha$, as
\begin{equation}
\label{Hzsmallalpha} 
	H^{2}(z)\simeq
	\bar{H}_\Lambda^{2}(z)+
	H_{100}^2\,
	\bar{\omega}_{c}
	\left[\left(\ln \left(\frac{1+z}{1+z_{*}}\right)-\frac{1}{3}\right)(1+z)^{3}+\frac{1}{3}+\ln(1+z_{*})\right]\alpha 
\,,
\end{equation}
where $\bar{H}_\Lambda(z)$ is given by Eq.~\eqref{barHz} with $\omega_c=\bar\omega_c$ fixed, i.e.,
\begin{equation}
\label{barbarHz}
	\bar{H}_\Lambda(z) = 
	H_{100} 
	\left[{\omega}_r\left\{(1+z)^4-1\right\}+ ({\omega}_b+\bar{\omega}_c)\left\{(1+z)^3-1\right\} + h^2\right]^{1/2} \,.
\end{equation}
Figure~\ref{fig:alpha-h_theta} illustrates the degeneracy between $\alpha$ and $h$ for a given value of the acoustic angular scale, $\theta_{*}$ in Eq.~\eqref{def:theta*}, where we fix the dark matter density at the time of recombination.

The right panel of Figure~\ref{fig:Cls_alpha} shows the effect of varying $\alpha$ while fixing both $\theta_{*}$ and $\omega_{c*}$. As expected this shows much smaller residual variations in the CMB spectra, mainly affecting the higher acoustic peaks. 
Nonetheless we shall see that this weak residual dependence on the interaction strength, due to the finite vacuum-matter interaction even before recombination, allows the CMB data on their own to break the degeneracy between $\alpha$ and $h$,
in particular placing bounds on the allowed range of the parameter, $\alpha$, in this model.

%

%

Finally we note that the redshift at matter-radiation equality in this model is given by the implicit equation
\begin{align}
\label{eq:a_eq_alpha}
{1+z_\mathrm{eq}}
= 
\frac{\omega_{b}+\omega_{c}(1+z_\mathrm{eq})^{\alpha}}{\omega_{r}}.
\end{align}
Given that $1+z_\mathrm{eq}\sim3400$, we find that for fixed values of the other cosmological parameters ($\omega_{r}$, $\omega_{b}$ and $\omega_{c}$) a positive $\alpha$ yields a larger numerator on the right-hand-side of Eq.~\eqref{eq:a_eq_alpha}, implying that matter-radiation equality occurs at a higher redshift than in the non-interacting case. This leads to a turnover in the matter power spectrum at smaller scales, and hence more power in the matter power spectrum for positive $\alpha$. This is demonstrated, for example, by a higher value for the mass variance, $\sigma_8^2$, on $8$~Mpc scales. If we again assume that CMB constraints will fix the matter density at the time of recombination according to Eq.~\eqref{fixomegacalpha} then 
\begin{align}
\label{eq:a_eq_alphafixomegac}
{1+z_\mathrm{eq}}
= 
\frac{\omega_{b}+\bar\omega_{c}[(1+z_\mathrm{eq})/(1+z_*)]^{\alpha}}{\omega_{r}}
\simeq \frac{\omega_{b}+\bar\omega_{c}[1+1.1\alpha]}{\omega_{r}}
\,,
\end{align}
where the final approximation holds for $\alpha\ll1$. Thus the effect of varying $\alpha$ on the matter power spectrum, though still present, is less pronounced once we apply CMB constraints on the matter density at recombination.

\subsection{Model: $Q=\beta HV$}

\begin{figure}
\centering
\includegraphics[width=\textwidth]{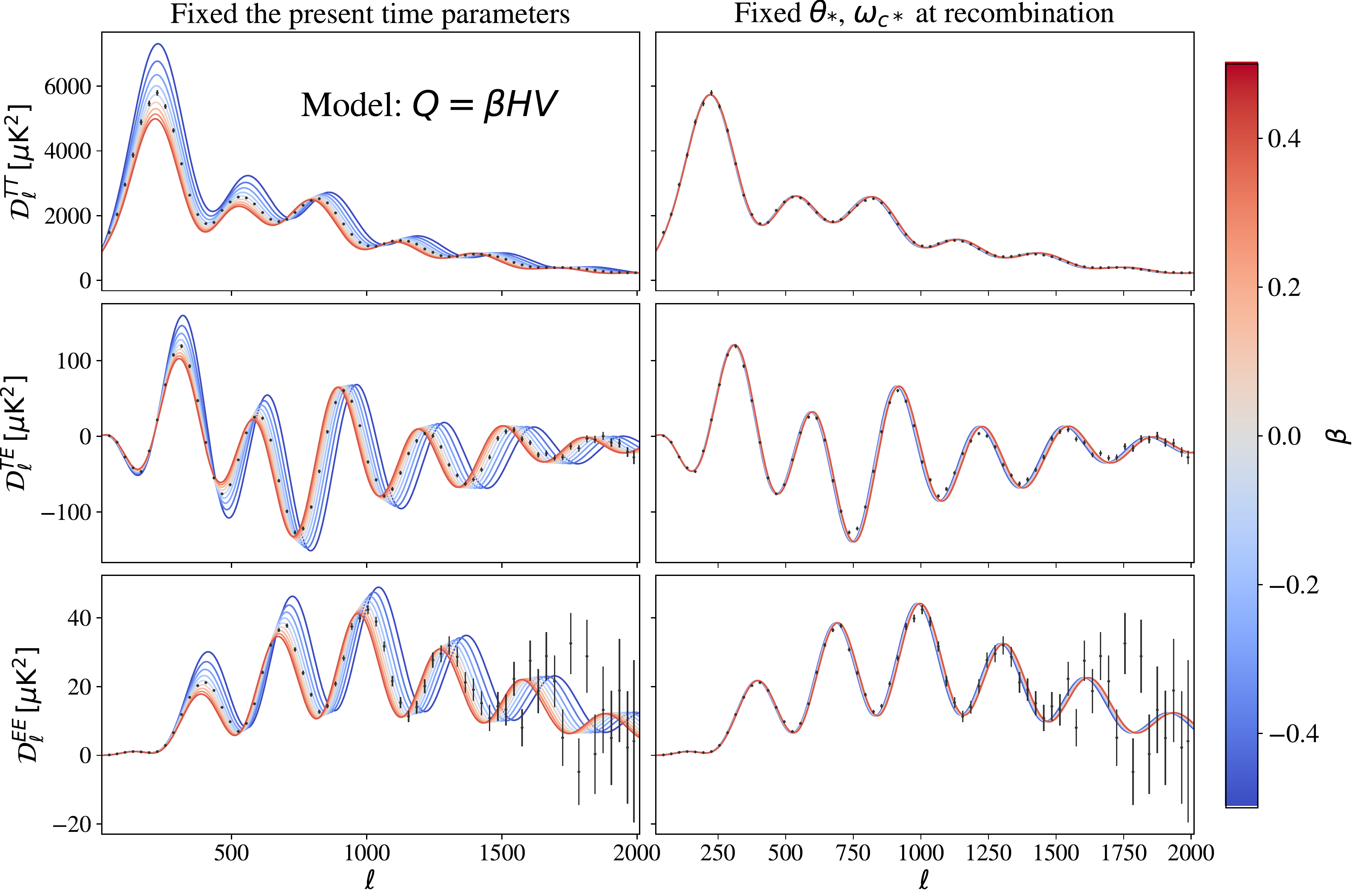}
\caption{\label{fig:Cls_beta}The plots on the left panel shows variations on CMB spectra in different value of $\beta$ with fixed present time parameters. The right panel shows tiny variance of the spectra with fixing the acoustic angle and physical dark mater density parameter at recombination and allowing $H_0$ and $\Omega_{c,0}$ to vary.}
\end{figure}

As previously noted, the heights of the acoustic peaks in the CMB angular power spectrum will constrain the dark matter density at recombination.
This is related to the present dark matter density by Eq.~\eqref{rhocbeta},
%
%
and since at recombination $z_*\gg1$, we find (for $\beta>-3$)
%
\begin{equation}
\label{omegac*beta}
	\omega_{c*}
	\simeq \left[ \omega_{c} + \frac{\beta}{\beta+3} \omega_V \right] (1+z_*)^3
	\,,
\end{equation}
where we have defined $\omega_V\equiv h^2-\omega_r-\omega_b-\omega_c$.
Since the energy transfer is proportional to the vacuum energy density, the interaction only becomes relevant once the vacuum energy becomes significant at late times for $\beta>-3$. As a result, at early times, and in particular around the time of recombination ($z_*\gg1$), the cold dark matter is effectively non-interacting and hence $\omega_{c*}\propto (1+z_*)^3$. For fixed value of the present matter density, $\omega_c$, we see that the matter density at recombination, $\omega_{c*}$, will be larger for positive $\beta$ and therefore the radiation driving will be weaker, reducing the height of the CMB acoustic peaks. This is illustrated in the left-hand panel of Figure~\ref{fig:Cls_beta}.

The expansion history, $H(z)$, is given in Eq.~\eqref{Hzbetaomegas} for this model as a function $\beta$, $h$ and the present-day energy densities.
%
For sufficiently small $\beta$ we can write this as
\begin{equation}
\label{Hzbetasmall}
    H^2(z) = {H}_\Lambda^2(z) +
    H_{100}^2\, \omega_V \left[ \frac{1}{3}(1+z)^3 - \frac{1}{3} - \ln(1+z) \right] \beta \,.
\end{equation}
Using Eq.~\eqref{omegacLCDM} to fix the dark matter density at recombination, $\omega_{c*}$ in Eq.~\eqref{omegac*beta}, to match the best-fit value in the $\Lambda$CDM model obtained for the present day density, $\bar\omega_c$,
%
%
%
we obtain the simple relation (for $|\beta|\ll1$)
\begin{equation}
\label{fixomegacsmallbeta}
    \bar\omega_c \simeq \omega_c + 
    \frac{\beta}{3}\omega_V
\,.
\end{equation}
Substituting this expression for $\omega_c$ into Eq.~\eqref{barHz} for $H_\Lambda(z)$, we can write Eq.~\eqref{Hzbetasmall} as
\begin{equation}
\label{Hzbetasmall2}
H^2(z) \simeq \bar{H}_\Lambda^2(z) - H_{100}^2\,\omega_V \left[\ln(1+z)\right] \beta
\,,
\end{equation}
where $\bar{H}_\Lambda(z)$, defined in Eq.~\eqref{barbarHz}, is the $\Lambda$CDM expansion history for $\omega_c=\bar\omega_c$.
This leads to a degeneracy between $\beta$ and $h$ in the the angular diameter distance to the surface of last-scattering, $D_{A*}$ in Eq.~\eqref{def:dA_to_rec}, and hence the angular acoustic scale, $\theta_*$ given in Eq.~\eqref{def:theta_s}, even after fixing the dark matter density at the time of recombination, which effectively fixes sound horizon, $r_*$ in Eq.~\eqref{eq:rs_rec}, if $\omega_b$ and $\omega_r$ are known.

\begin{figure}
\centering
\includegraphics[width=0.7\textwidth]{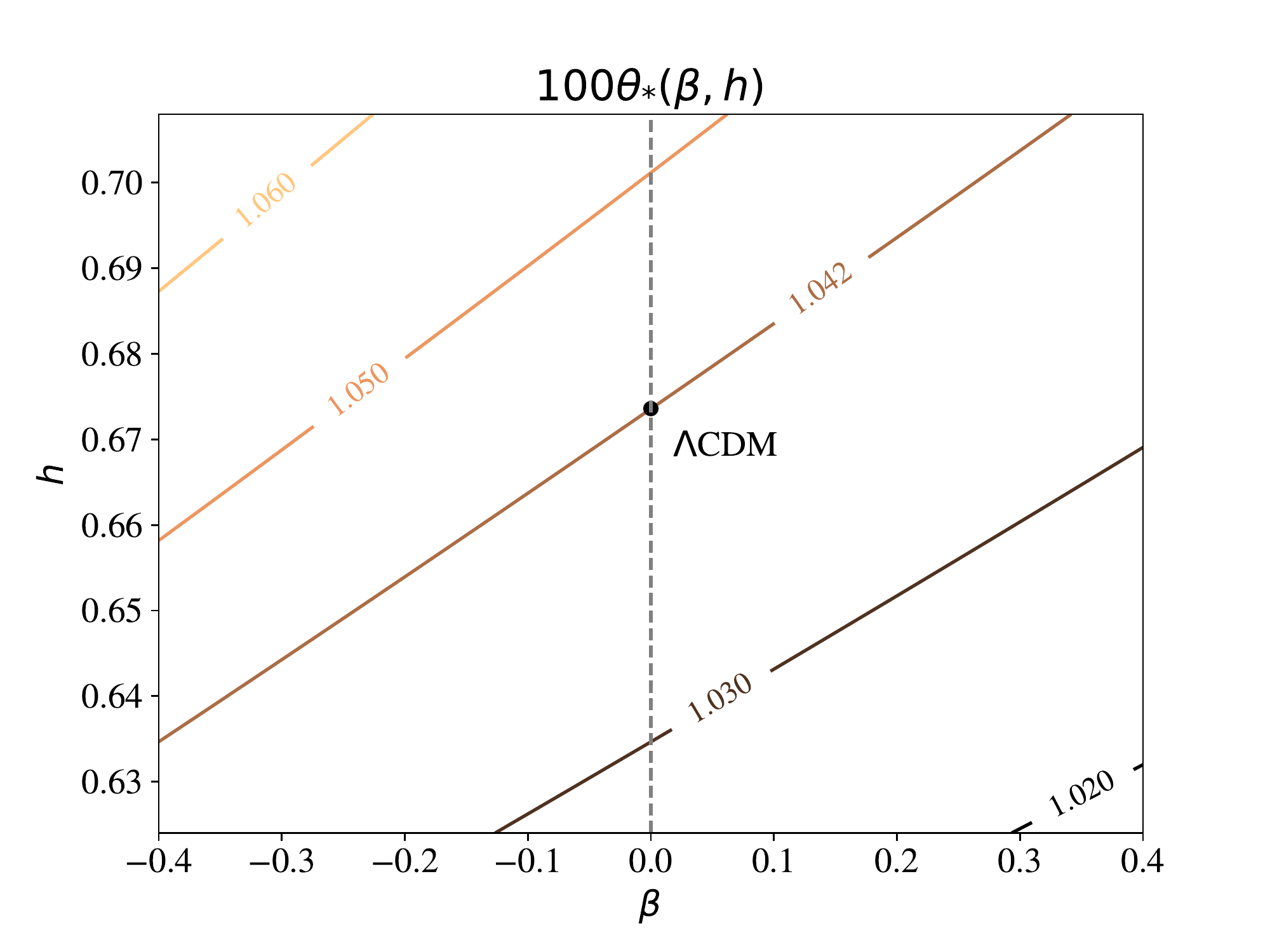}
\caption{\label{fig:beta-h_theta}Contour plot showing the angular acoustic scale, $\theta_{*}$, for different values of $h$ and $\beta$ while fixing the dark matter density at the time of recombination according to Eq.~\eqref{fixomegacsmallbeta}. The middle diagonal line corresponds to the best-fit value of $\theta_*$ in the $\Lambda$CDM model.}
\end{figure}

This degeneracy for $\theta_*$ is illustrated in Figure~\ref{fig:beta-h_theta}. Note however the large range shown for $\beta$ on the x-axis in Figure~\ref{fig:beta-h_theta} compared with the much narrower range shown for $\alpha$ in the corresponding Figure~\ref{fig:alpha-h_theta}. The expansion history, $H(z)$, and hence the angular acoustic scale, $\theta_*$, is much less sensitive at high redshift to $\beta$, since the additional $\beta$-dependent term in Eq.~\eqref{Hzbetasmall2} is only proportional to $\ln(1+z)$, rather than $(1+z)^3$ which appears in the $\alpha$-dependent term in Eq.~\eqref{Hzsmallalpha}.

%
%


%

The right panel of Figure~\ref{fig:Cls_beta} shows that if we fix both the height of the CMB acoustic peaks and the angular acoustic scale (by fixing the angular diameter distance to the CMB) then the residual variation of the CMB power spectra is tiny even for $|\beta|\sim0.5$. 
As a result the observed CMB angular power spectra alone cannot break the degeneracy between $\beta$ and $h$, in contrast to the previous model with $\alpha\neq0$.
This raises the possibility, for example, of resolving some of the tensions between the value of the Hubble constant inferred from CMB data in the $\Lambda$CDM model and the value measured by low-redshift probes. We need to include additional astrophysical data-sets to break this degeneracy. 

For small $\beta$ the redshift at matter-radiation equality can be written in the implicit form
\begin{equation}
\label{eq:a_eq_beta}
1+z_\mathrm{eq} 
\simeq \frac{\omega_b+\omega_c+
(\beta/3)
\omega_{V}}{\omega_{r}}
\,.
\end{equation}
%
The radiation-matter equality takes place earlier (at higher redshift $z_\mathrm{eq}$) if $\beta$ is positive, for fixed $\omega_b$ and $\omega_c$. 
However, if we fix the dark matter density at recombination according to Eq.~\eqref{fixomegacsmallbeta}, then we recover the standard expression
\begin{equation}
\label{eq:a_eq_beta2}
1+z_\mathrm{eq} \simeq \frac{\omega_b+\bar\omega_c}{\omega_{r}}\,.
\end{equation}
This suggests that if CMB observations fix the dark matter density at recombination, then the redshift of matter-radiation equality is also fixed in this model and hence so is the turnover in the matter power spectrum.


\section{Redshift-space distortions}
\label{sec:rsd}


If we use only the observed redshift to assign a radial distance to galaxies using Hubble's law, then this gives the coordinate position of the galaxies in {\em redshift space}
\begin{equation}
r_s = H^{-1} v = r + H^{-1} v_p \,.
\end{equation} 
The peculiar velocity along the line of sight, $v_p$, with respect to the Hubble flow thus leads to a distortion in redshift space with respect to the Newtonian {\em real space} distance, $r$.
%
This leads to a quadrupole anisotropy in the large-scale galaxy distribution in spectroscopic galaxy surveys, where the radial distance is inferred from the line-of-sight velocity~\cite{Percival:2008sh,Song:2008qt}.
The quadrupole moment of the galaxy distribution probes the matter-velocity cross-correlation~\cite{Planck:2015fie}.

%
%

In $\Lambda$CDM cosmology with, for non-interacting dark matter ($Q=0$), the Eulerian velocity divergence is related to the growth of the comoving density contrast via\footnote{Comparing to Eq.~\eqref{comsynevol:delta} for $Q=0$, we have the relation $\vb{v}_p=\delr\vartheta$.}
\begin{equation}
\label{LCDMtheta}
\mathcal{V} \equiv
- \frac{\divr{\vb{v}}_p}{H} = H^{-1} \dot{\delta}_m = f\delta_m\,,
\end{equation}
where $\delta(\mathbf{x},t)=\delta_0(\mathbf{x})D_{+}(t)$ with a time-independent $\delta_0(\mathbf{x})$, and the growth rate $f$ is related to the growth factor $D_{+}$ as follow
\begin{equation}
\label{deff}
f \equiv \frac{\dot{D}_+}{HD_+}
\,.
\end{equation} 
For non-interacting matter the redshift-space distortions can be related to the variance of the growth of the matter density contrast via Eq.~\eqref{LCDMtheta}, and is conventionally written as a function of redshift
%
\begin{equation}
\label{LCDMfsigma8}
    \qty[\frac{\langle
     \mathcal{V}\delta_m\rangle}{\langle\delta_m^2\rangle^{1/2}}]_{8h^{-1}{\rm Mpc}}
    \equiv f(z) \sigma_8(z)\,.
\end{equation}
%
%
$\sigma_8(z)$ is the root-mean square (RMS) of the density fluctuation in a sphere of comoving radius $8~h^{-1}\mathrm{Mpc}$ defined as
\begin{equation}
\label{eq:sigma_eight}
\sigma_8(z) \equiv \expval{\delta_m^2}_{8h^{-1}{\rm Mpc}}^{1/2} = \qty[\frac{1}{2\pi^2}\int_0^\infty k^2 P_L(k,z)\abs{W_8(k)}^2\dd{k}]^{1/2} \,,
\end{equation}
where 
\begin{equation}
P_L(k,z)=P_L(k)\qty[\frac{\delta_m(z)}{\delta_m(0)}]^2 \,,
\end{equation}
is the linear matter power spectrum in Fourier space, and $W_8(k)$ is the Fourier transform of the a spherical top-hat window function of width $R=8\,h^{-1}{\rm Mpc}$. Therefore the redshift-dependence of the RMS density fluctuation can be written as $\sigma_8(z)=\sigma_8(0)D_{+}(z)$, where $\sigma_8(0)$ is the present time value and $D_{+}(0)$ is normalised to unity. 


In the presence of interacting dark matter the continuity equation, Eq.~\eqref{LCDMtheta}, is modified to
become (see Eq.~\eqref{comsynevol:delta})
\begin{equation}
\dot\delta_m + \divr{\vb{v}}_p = \frac{Q}{\rho_m} \delta_m
\,.
\end{equation}
Thus the redshift-space distortions are related to the variance in the density contrast via a modified relation
%
\begin{equation}
\label{intfsigma8}
    \qty[\frac{\langle
     \mathcal{V}\delta_m\rangle}{\langle\delta_m^2\rangle^{1/2}}]_{8h^{-1}{\rm Mpc}} = f_\mathrm{rsd}(z) \sigma_8(z)\,,
\end{equation}
%
%
where~\cite{Wang:2015wga,Borges:2017jvi}
\begin{equation}
\label{deffrsd}
f_\mathrm{rsd} = f - \frac{Q}{H\rho_m} \,.
\end{equation}
Only when $Q/H\rho_m\to0$ we recover the non-interacting result, $f_\mathrm{rsd}\to f$, given in Eq.~\eqref{LCDMfsigma8}. Thus care must be taken when relating redshift-space distortions to $f\sigma_8$ and the growth of structure in models with interacting dark matter. 

In the following we will use Eq.~\eqref{intfsigma8} to compare our theoretical models with observational bounds on redshift-space distortions, where $f_\mathrm{rsd}$ is related to the growth rate $f$ via Eq.~\eqref{deffrsd}.
The growth rate, $f$, is given by 
Eq.~\eqref{deff}, where $D_{+}$ is the growth factor for matter. The growth factor itself is given by the solution to Eq.~\eqref{eq:growth_delta_m}, which includes the interaction.
%
%


\section{Observational constraints}
\label{sec:observations}

\subsection{Likelihoods}

We consider constraints on the linear interaction model parameters coming from cosmic microwave background temperature anisotropies and polarisation data, together with type-Ia supernova data, and baryon acoustic oscillation and redshift-space distortions, observed in galaxies surveys of large-scale structure. We give a short description of the data sets in the following:

\begin{itemize}
    \item \textbf{Cosmic Microwave Background:} we use likelihoods from the Planck 2018 data release, using the Planck TT power spectrum with high and low multipoles in the range $\ell\geq30$ and $2\leq\ell<29$ respectively. 
    We also considered the $E$-mode polarisation observation of CMB by using Planck TT, TE, EE which is the combined Planck TT, Planck TE, and Planck EE power spectra at $\ell>29$, together with low-$\ell$ Planck TT and EE spectra for multipoles in the range $2\leq\ell<30$
    \cite{Aghanim:2019ame}.
\item \textbf{Type-Ia Supernova:} we used the dataset from 1048 type Ia supernovae (SNe\,Ia) in the redshift range $0.01<z<2.3$ called the \textit{Pantheon Sample}~\cite{Scolnic:2017caz}. 
This SN\,Ia dataset provides the logarithm of the overall flux normalisation, $m_B$, which is related to the distance modulus, $\mu=m_B-M$, where $M$ is the unknown absolute magnitude. However, in the MCMC analysis, the absolute magnitude $M$ can be treated as a nuisance parameter. The distance modulus can be written as
\begin{equation}
	\mu 
	= 5\log_{10}\left(\frac{d_{L}}{1\,\mathrm{Mpc}}\right)+25,
\end{equation} 
where $d_L$ is the luminosity distance which depends on cosmological parameters. For a flat FLRW cosmology, the luminosity distance reads
\begin{equation}
d_L(z) = {(1+z)c}\int_{0}^{z}\frac{\dd{z'}}{H(z')},
\end{equation}
where the Hubble function $H(z)$ contains $H_0$ and the interaction parameters as discussed in subsections~\ref{ssec:a_model} and~\ref{ssec:b_model}.

%

    %
    
    \item \textbf{Baryon Acoustic Oscillations and Redshift Space Distortion:} we used five data points from the combined measurements of BAO and RSD data provided from the SDSS database\footnote{\url{https://svn.sdss.org/public/data/eboss/DR16cosmo/tags/v1_0_1/likelihoods/}}. 
    We utilised the BAO-RSD likelihoods from the first two BOSS DR12 likelihood results, with effective redshift $z_\mathrm{eff}=0.38$ and $z_\mathrm{eff}=0.51$, which is extracted from~\cite{Alam:2016hwk}. We used the likelihood for $z_\mathrm{eff}=0.70$ from the combined eBOSS and BOSS results from the measurement of the BAO and growth rate of structure of the Luminous Red Galaxy sample (LRG) \cite{Bautista:2020ahg,Gil-Marin:2020bct}. For high-redshift BAO-RSD, we adopted the BAO-RSD likelihood from the eBOSS quasar sample (QSO) with effective redshift $z_\mathrm{eff}=1.48$~\cite{Hou:2020rse,Neveux:2020voa}. All of these likelihood data provide measurements of\footnote{In BAO papers, the notation $D_M$ is commonly used instead of $D_A$} $D_A(z)/r_d$, $D_H(z)/r_d$, and $f\sigma_8$, where $D_A(z)$ stands for the comoving angular diameter distance. In a flat FLRW at redshift $z$, $D_A(z)$ can be expressed as
\begin{equation}
	D_A(z) = \frac{c}{H_0}\int_{0}^{z}\frac{\dd{z'}}{E(z')},
\end{equation} 
with $E(z)\equiv H(z)/H_0$. $D_H(z)\equiv c/H(z)$ is the Hubble distance. The variable $r_d$ is the comoving sound horizon at the baryon drag epoch can be written as
\begin{equation}
r_d = \frac{1}{H_0}\int_{z_d}^{\infty}\frac{c_s(z)}{E(z)}\dd{z},
\end{equation}
where $z_d$ is the redshift at the drag epoch 
when the baryons decouple from the photons shortly after recombination. 

In the interaction model we the growth of structure modified to be $f\sigma_8\rightarrow f_\mathrm{rsd}\sigma_8$, where $f_\mathrm{rsd}$ and $\sigma_8$ are given by Eqs.~\eqref{deffrsd} and~\eqref{eq:sigma_eight}, respectively. 

For the effective redshift $z_\mathrm{eff}=0.15$, the likelihood was taken from the measurement of the clustering of the SDSS Main Galaxy Sample (MGS) \cite{Howlett:2014opa} which provides measurements of $f_\mathrm{rsd}\sigma_8$ and $D_V(z)/r_d$, where the spherically-averaged distance is given by
\begin{equation}
D_V(z) = \left[zD_A^2(z)D_H(z)\right]^{1/3}.
\end{equation}

The BAO-RSD datasets are summarised in the Table~\ref{Tab:bao_rsd_data_table} which is taken from Table~3 of~\cite{eBOSS:2020yzd}.
These data points are plotted in Figures~\ref{fig:D_alpha} and~\ref{fig:fsigma8_alpha} alongside the theoretical predictions using Eq.~(\ref{eq:Hz_alpha}) for $H(z)$ in the $\alpha$-model.
We show in Figures~\ref{fig:D_beta} and~\ref{fig:fsigma8_beta} the theoretical predictions using Eq.~(\ref{Hzbetaomegas}) for $H(z)$ in the $\beta$-model.



\begin{figure}
\includegraphics[width=\textwidth]{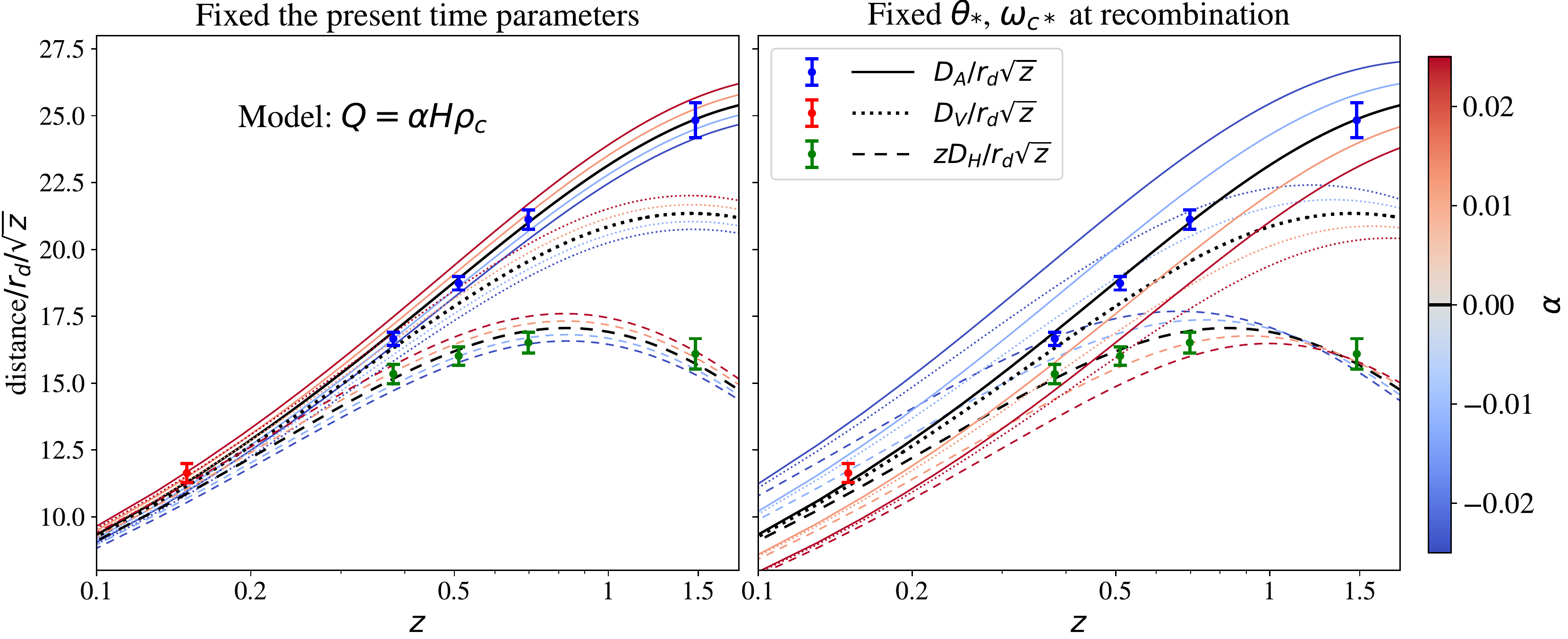}
\caption{\label{fig:D_alpha}The left panel shows the effect  of varying $\alpha$ on different distance measures ($D_A(z)$, $D_V(z)$ and $D_H(z)$) while holding present-day cosmological parameters ($\omega_c$ and $h$) fixed. The right panel shows the effect of varying $\alpha$ while fixing the CMB angular acoustic scale, $\theta_*$, and the physical dark matter density parameter at recombination, $\omega_{c*}$. The data points correspond to values listed in Table~\ref{Tab:bao_rsd_data_table}.}
\end{figure}

\begin{figure}
\includegraphics[width=\textwidth]{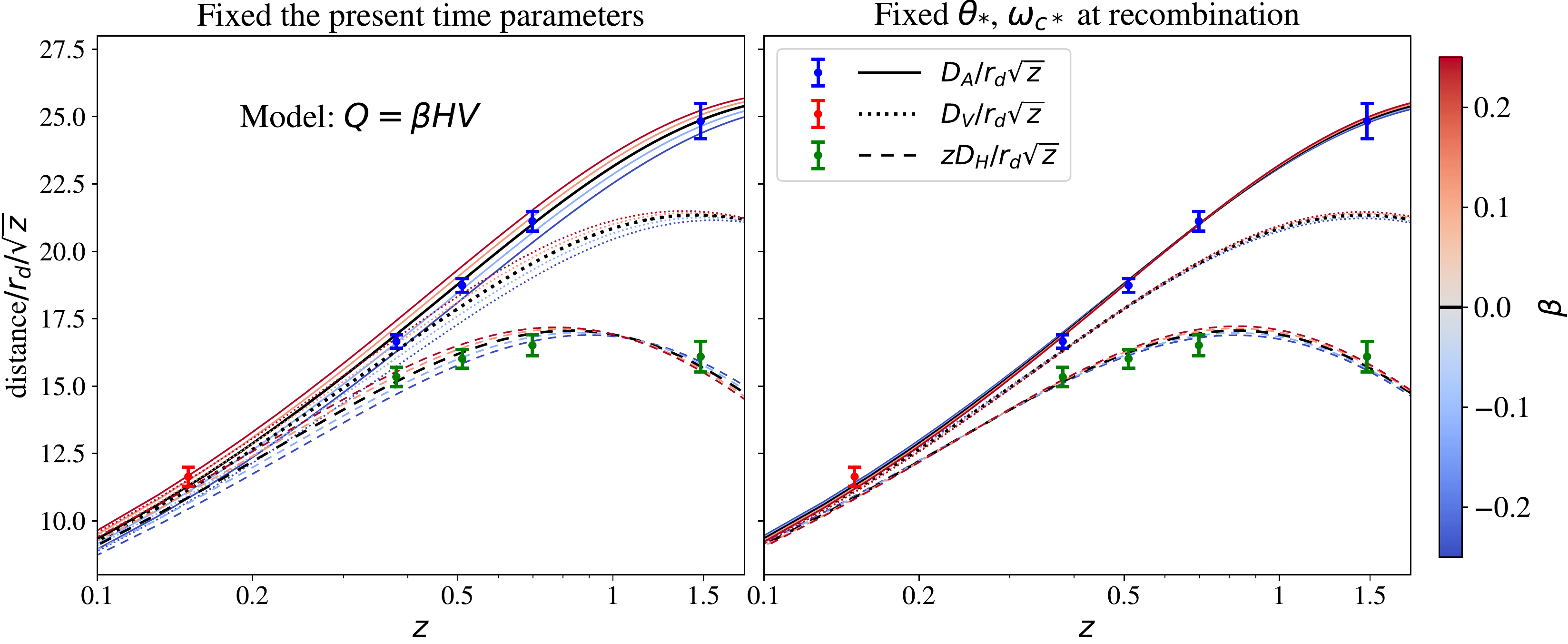}
\caption{\label{fig:D_beta}The left panel shows the effect  of varying $\beta$ on different distance measures ($D_A(z)$, $D_V(z)$ and $D_H(z)$) while holding present-day cosmological parameters ($\omega_c$ and $h$) fixed. The right panel shows the effect of varying $\beta$ while fixing the CMB angular acoustic scale, $\theta_*$, and the physical dark matter density parameter at recombination, $\omega_{c*}$. The data points correspond to values listed in Table~\ref{Tab:bao_rsd_data_table}.}
\end{figure}

\begin{figure}
\includegraphics[width=\textwidth]{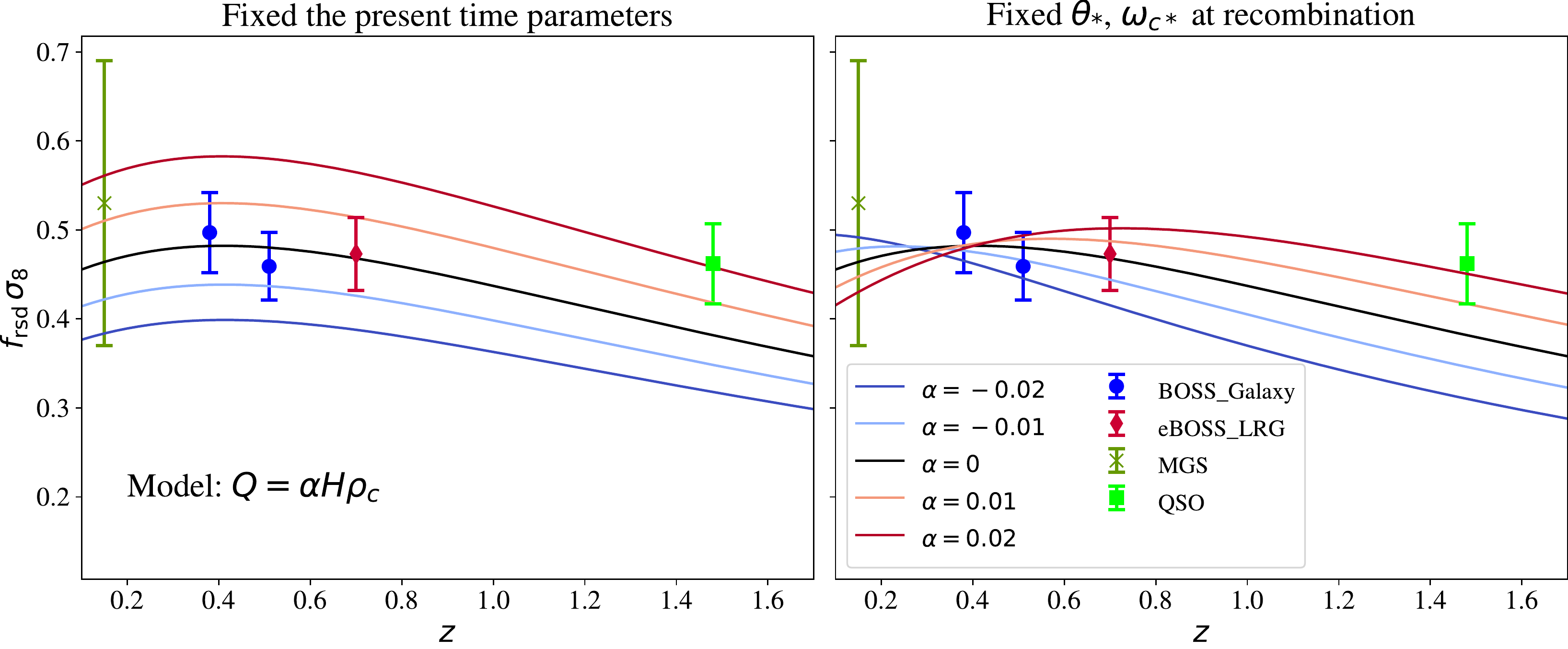}
\caption{\label{fig:fsigma8_alpha}The left panel shows as the effect on $f_\mathrm{rsd}\sigma_8(z)$ of varying $\alpha$ while holding present-day cosmological parameters ($\omega_c$ and $h$) fixed. The right panel shows $f_{\rm rsd}\sigma_8$ when varying $\alpha$ while instead fixing the CMB angular acoustic scale, $\theta_*$, and the physical dark matter density parameter at recombination, $\omega_{c*}$. The data points correspond to values of $f_{\rm rsd}\sigma_8$ listed in the final row of  Table~\ref{Tab:bao_rsd_data_table}.}
\end{figure}

\begin{figure}
\includegraphics[width=\textwidth]{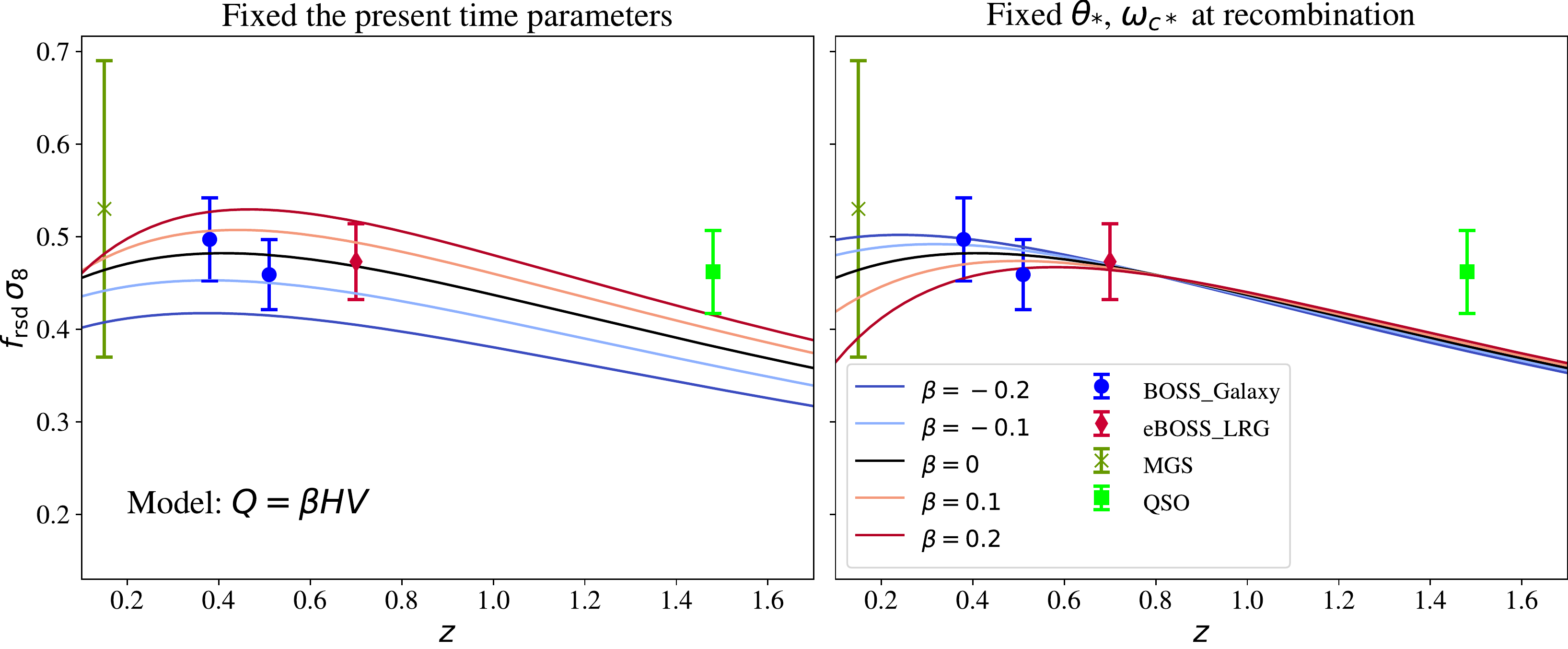}
\caption{\label{fig:fsigma8_beta}The left panel shows the effect on $f_\mathrm{rsd}\sigma_8(z)$ of varying $\beta$ while holding present-day cosmological parameters ($\omega_c$ and $h$) fixed. The right panel shows $f_{\rm rsd}\sigma_8$ when varying $\beta$ while instead fixing the CMB angular acoustic scale, $\theta_*$, and the physical dark matter density parameter at recombination, $\omega_{c*}$. The data points correspond to values listed in the final row of  Table~\ref{Tab:bao_rsd_data_table}.}
\end{figure}


\begin{table}
\caption{\label{Tab:bao_rsd_data_table}The likelihood data for each of the BAO-RSD measurements used in this paper taken from the Table~3 of~\cite{eBOSS:2020yzd}.}
\begin{tabular}{cccccc}
\hline\hline
$z_\mathrm{eff}$ & 0.15          & 0.38            & 0.51            & 0.70            & 1.48            \\
\hline 
$D_V(z)/r_{d}$   & $4.51\pm0.14$ &                 &                 &                 &                 \\
$D_A(z)/r_{d}$   &               & $10.27\pm0.15$  & $13.38\pm0.18$  & $17.65\pm0.30$  & $30.21\pm0.79$  \\
$D_H(z)/r_{d}$   &               & $24.89\pm0.58$  & $22.43\pm0.48$  & $19.78\pm0.46$  & $13.23\pm0.47$  \\
$f\sigma_8(z)$   & $0.53\pm0.16$ & $0.497\pm0.045$ & $0.459\pm0.038$ & $0.473\pm0.041$ & $0.462\pm0.045$ \\
\hline
\end{tabular}

\end{table}






\end{itemize}

\subsection{Implementation}

In order to test the models against observations, we modified the Einstein-Boltzmann code \texttt{CLASS}\footnote{\url{https://github.com/lesgourg/class_public}} \cite{2011arXiv1104.2932L, 2011JCAP07034B} to allow for an interaction between the dark matter and the vacuum energy, according to the interaction model \eqref{eq:linear_model_Q}. 
%
We run a Markov-chain Monte Carlo (MCMC) analysis using \texttt{MontePython}\footnote{\url{https://github.com/brinckmann/montepython_public}} \cite{Audren:2012wb,Brinckmann:2018cvx}, with a Metropolis-Hasting algorithm to sample the multi-parameter distribution, verifying convergence using the Gelman-Rubin criterion.

We introduce the analytic solution of the background continuity equations, Eq.~\eqref{eq:idm_iv_fld_Q}, into the background module of \texttt{CLASS} with boundary conditions set in terms of the present day densities. 
We implemented the evolution of the linear density contrast of cold dark matter perturbations in the comoving synchronous gauge, given in Eq.~\eqref{synevol:deltac}. 
We solve the linear growth factor given by the growing mode solution of the linear growth equation including the interaction terms, Eq.~\eqref{eq:growth_delta_m}, 
together with the modified growth function Eq.~\eqref{deffrsd} numerically.

\subsection{Results}
\label{ssec:results}

\begin{figure}[h!]
\centering
\includegraphics[width=0.8\textwidth]{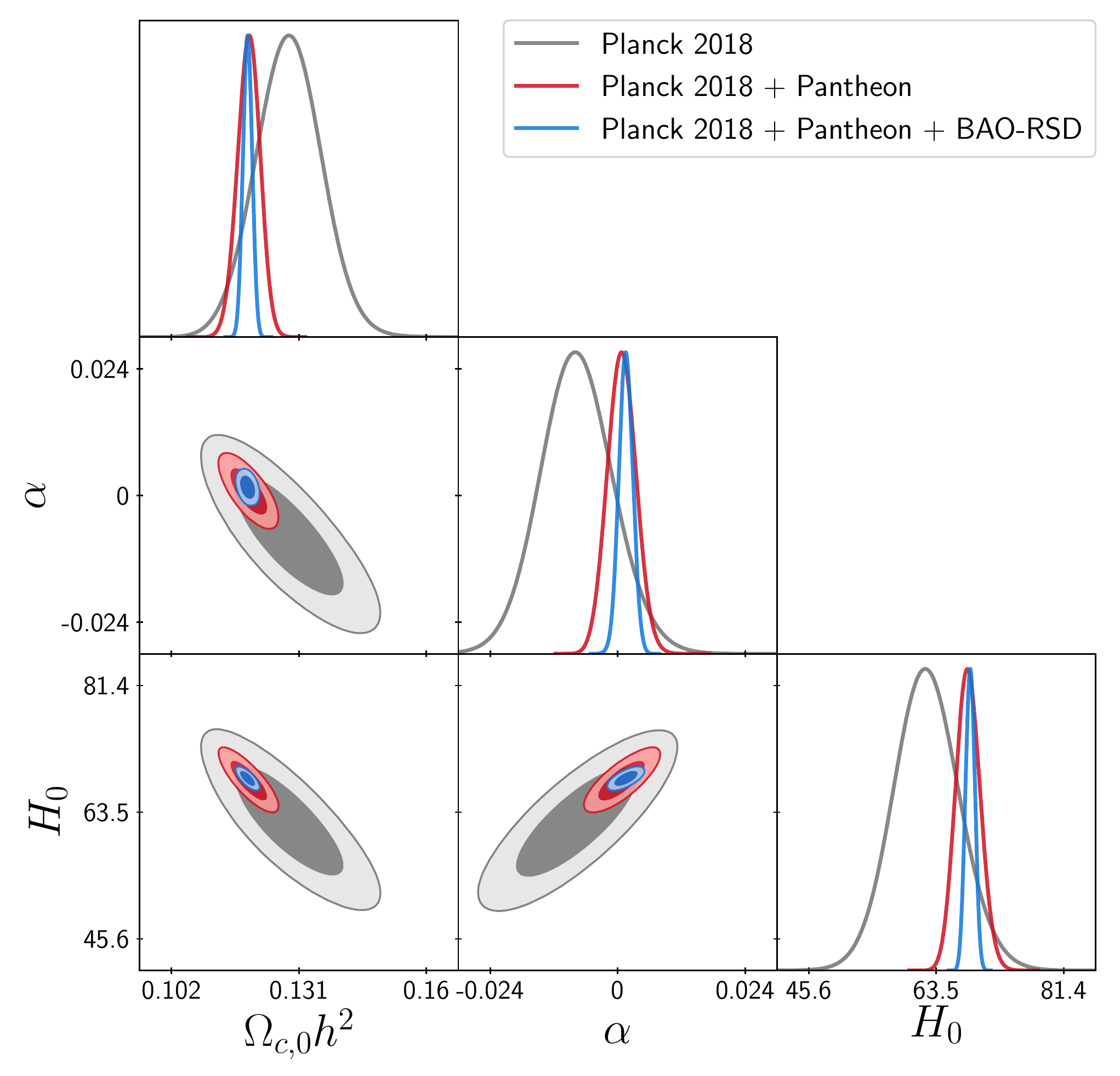}
\caption{\label{fig:a_com_P18_SNIa_BAO-RSD}
1D and 2D marginal distributions for $\alpha$, $H_0$ and $\Omega_{c,0}h^2$ in the $\alpha$-model (with $\beta=0$) showing the distributions from Planck 2018 CMB data (TT, TE and EE and lowE, in grey), plus Pantheon supernova data (in red) and BAO-RSD data (in blue).}
\end{figure}

We show in Figure~\ref{fig:a_com_P18_SNIa_BAO-RSD} the observational constraints on the interaction model $Q=\alpha H\rho_c$ with $\beta=0$. 

Using the Planck data alone we see an approximate degeneracy between the 3 parameters $\alpha$, $H_0$ and $\Omega_{c,0}h^2$. 
As discussed in section~\ref{sec:cmb_aniso}, we see that the CMB data fix the dark matter density at recombination, $\omega_{c*}$ in Eq.~(\ref{omegac*}), which leaves a degeneracy between $\alpha$ and $\Omega_{c,0}h^2$, given for small $\alpha$ by Eq.~(\ref{degen:omegac-alpha}). Substituting the $\Lambda$CDM central value for $\bar\omega_c\simeq0.12$ (see Table~\ref{tab:P18_six_param}) into Eq.~(\ref{degen:omegac-alpha}), gives
\begin{equation}
    \Omega_{c,0}h^2 \simeq 0.12 - 0.84\alpha \,.
\end{equation}
This matches the approximate degeneracy between $\alpha$ and $\Omega_{c,0}h^2$ shown in Figure~\ref{fig:a_com_P18_SNIa_BAO-RSD}.
In addition the CMB data tightly constrain the acoustic angular scale at recombination, $\theta_*$. Thus the degeneracy between $\alpha$ and $H_0$ seen in Figure~\ref{fig:a_com_P18_SNIa_BAO-RSD} follows the contours for constant $\theta_*$ plotted in Figure~\ref{fig:alpha-h_theta}. 

Despite these approximate degeneracies, the CMB data alone are still sufficient to bound $\alpha=-0.0077^{+0.0069}_{-0.0071}$ (and hence $H_0$ and  $\Omega_{c,0}h^2$), due to the residual dependence on $\alpha$ of the second and higher acoustic peaks in the temperature and polarisation power spectra, shown in the right-hand panel of Figure \ref{fig:Cls_alpha}.
The parameter constraints coming from Planck data are shown in the left-hand column in Table~\ref{table:alphaparams}.

\begin{table}
\caption{\label{table:alphaparams}
Parameter constraints in the interaction model $Q=\alpha H\rho_c$ (where $\beta=0$). Parameters below the dashed line are derived parameters.}
\def\arraystretch{1.7}
\begin{tabularx}{\textwidth}{Xccc} 
\hline\hline
Parameters                               & {TT,TE,EE+lowE}               & {+Pantheon}                  & {+ Pantheon + BAO-RSD}         \\ \hline
$100\,\Omega_{b,0} h^2$\dotfill & $2.242\pm{0.016}$             & $2.237\pm{0.016}$            & $2.233\pm{0.016}$      \\
$100\,\theta_{*}$  &  $1.04173\pm{0.00034}$  & $1.04193\pm{0.00030}$  & $1.04191\pm{0.00029}$ \\
$\tau$ \dotfill & $0.0523_{-0.0082}^{+0.0076}$ & $0.0546_{-0.0081}^{+0.0076}$ & $0.0559_{-0.0083}^{+0.0075}$ \\
$\Omega_{c,0} h^2$   \dotfill & $0.1282_{-0.0079}^{+0.0073}$  & $0.1191_{-0.0025}^{+0.0026}$ & $0.1189\pm{0.0010}$   \\
$\mathrm{ln}(10^{10}A_s)$       \dotfill & $3.038\pm{0.017}$             & $3.045_{-0.017}^{+0.016}$    & $3.049_{-0.017}^{+0.016}$      \\
$n_{s}$                         \dotfill & $0.9585\pm{0.0069}$           & $0.9656\pm{0.0045}$          & $0.9651_{-0.0041}^{+0.0039}$    \\
$\alpha$                        \dotfill & $-0.0077_{-0.0071}^{+0.0069}$ & $0.0008_{-0.0028}^{+0.0026}$ & $0.0016\pm{0.0014}$ \\
\hdashline
$\Omega_{V,0}$         \dotfill & $0.599_{-0.064}^{+0.097}$     & $0.692_{-0.020}^{+0.022}$    & $0.6954_{-0.0076}^{+0.0080}$    \\
$H_{0}$                         \dotfill & $62.2_{-4.9}^{+4.7}$          & $68.0_{-1.8}^{+1.7}$         & $68.27\pm{0.65}$        \\
$\sigma_{8}$                    \dotfill & $0.759_{-0.049}^{+0.047}$     & $0.815_{-0.021}^{+0.020}$    & $0.824_{-0.014}^{+0.013}$     \\
$S_{8}$                    \dotfill      & $0.8658_{-0.0040}^{+0.0035}$  & $0.8237\pm{0.0017}$          & $0.828\pm{0.013}$     \\
\hline 
\end{tabularx}

\end{table}

The parameter bounds are considerably improved by the addition of low-redshift data, first by the inclusion of the Pantheon compilation of supernovae, and then further by the BAO-RSD data from the SDSS survey, as shown in Figure~\ref{fig:a_com_P18_SNIa_BAO-RSD}. In combination this tightens the bound on the dark matter density by a factor of six, leading to a much tighter bound on $\alpha=0.0016\pm0.0014$, see the right-hand column in Table~\ref{table:alphaparams}.
The eBOSS data points at $z=1.48$, coming from observations of quasars and included in the BAO-RSD data in Table~\ref{Tab:bao_rsd_data_table}, play an important role in in the final constraint on $\alpha$, as can be most clearly seen in the right-hand panel of Figure~\ref{fig:fsigma8_alpha}. Without this quasar data we would obtain a lower central value $\alpha=0.0005\pm0.0014$.

\begin{figure}[h!]
\centering
\includegraphics[width=0.8\textwidth]{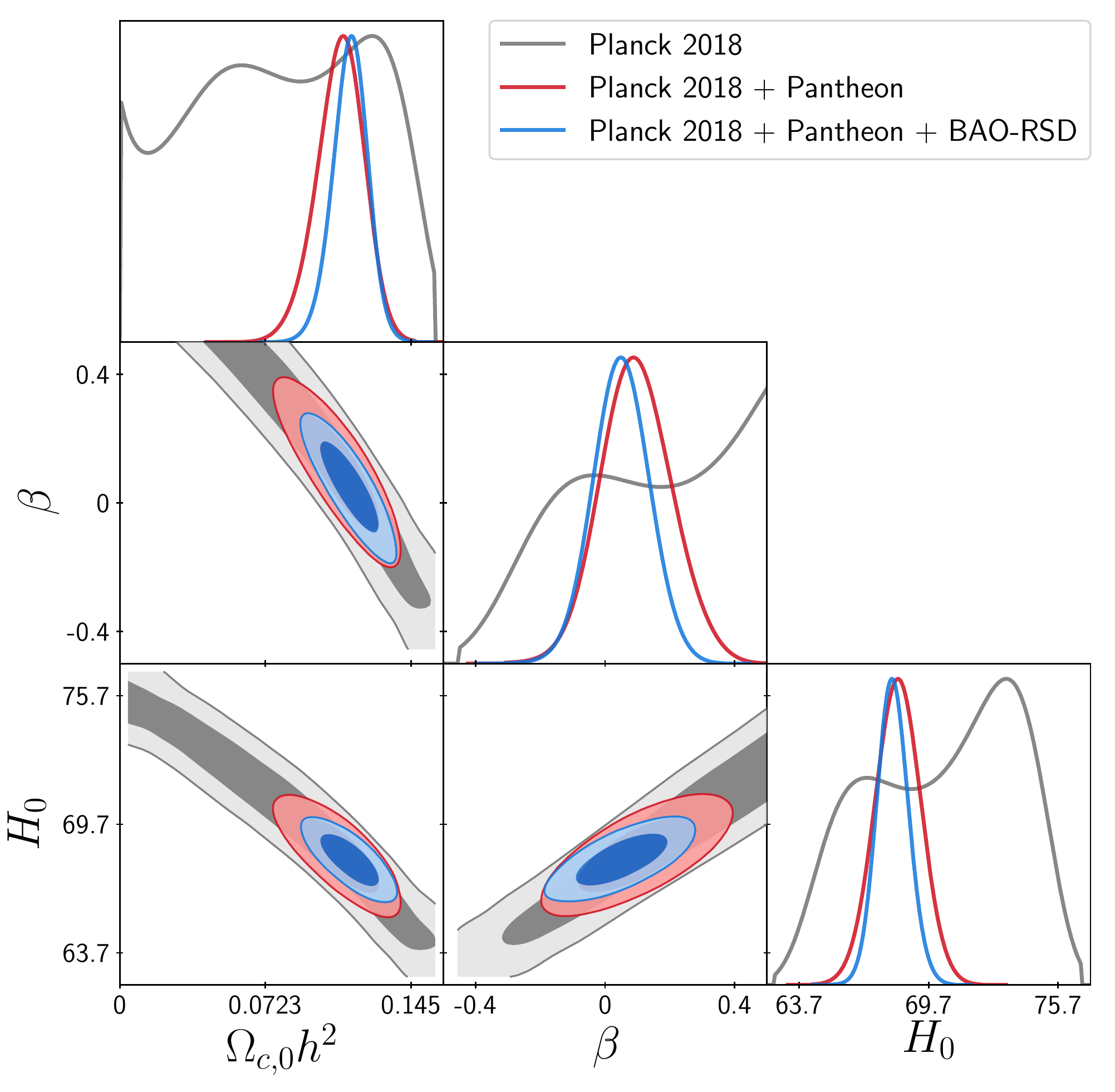}
\caption{\label{fig:b_com_P18_SNIa_BAO-RSD}
1D and 2D marginal distributions for $\alpha$, $H_0$ and $\Omega_{c,0}h^2$ in the $\beta$-model (with $\alpha=0$) showing the distributions from Planck 2018 CMB data (TT, TE and EE and lowE, in grey), plus Pantheon supernova data (in red) and BAO-RSD data (in blue).}
\end{figure}

We show the observational constraints on parameters in the interaction model $Q=\beta HV$, with $\alpha$ set to zero, in Figure~\ref{fig:b_com_P18_SNIa_BAO-RSD}.
%
There is a clear degeneracy between the three parameters $\beta$, $H_0$ and $\Omega_{c,0}h^2$ if we use only the CMB data, which leaves the interaction parameter $\beta$ very poorly constrained (hence we do not display the 1D marginal distributions for the parameters in this case) since in this case, the energy transfer between dark matter and the vacuum is negligible at early times. The dark mattter is effectively non-interacting around the time of recombination, and once we fix the dark matter density at the time of recombination in this model, $\omega_{c*}$ in Eq.~\eqref{omegac*beta}, then the CMB power spectra are indistinguishable from $\Lambda$CDM. Substituting the $\Lambda$CDM central values for $\bar\omega_c\simeq 0.12$ and $\bar\omega_V\simeq0.31$ (cf. Table~\ref{tab:P18_six_param}) into Eq.~(\ref{fixomegacsmallbeta}) we obtain
\begin{equation}
    \Omega_{c,0}h^2 \simeq 0.12 - 0.10\beta \,.
\end{equation}
On the other hand, the degeneracy between $\beta$ and $H_0$ closely follows the contours shown in Figure~\ref{fig:beta-h_theta} corresponding to a fixed angular acoustic scale, $100\theta_*\simeq 1.042$.

This degeneracy is only broken by including low-redshift data. The Pantheon data constrain the background evolution, $H(z)$. For small $\beta$ (and fixed dark matter density at recombination) this is given by Eq.~(\ref{Hzbetasmall2}). Hence we see that constraints on $H(z)$ provide an independent constraint on $\omega_V$, $\beta$ which breaks the degeneracy between $\beta$, $H_0$ and $\Omega_{c,0}h^2$ leading to the bound $\beta=0.10\pm0.11$, as presented in Table~\ref{table:betaparams}. Inclusion of the BAO-RSD data further constrains not just the background evolution but also the growth of structure, leading to the tighter bound $\beta=0.051^{+0.089}_{-0.087}$. We can see in the right-hand panel of Figure~\ref{fig:D_beta} that once the angular acoustic scale and dark matter density at recombination are fixed then there is very little variation with $\beta$ of the background evolution and thus the distance measures $D_A(z)$ or $D_H(z)$. However $\beta\neq0$ does affect the growth of structure and $f_{\rm rsd}\sigma_8$ shown in Figure~\ref{fig:fsigma8_beta}. Again we find that the eBOSS quasar data at $z=1.48$ plays an important role, and without this data we would find a lower central value $\beta=0.012_{-0.089}^{+0.093}$.

\begin{table}
\caption{\label{table:betaparams}Parameter constraints in the interaction model $Q=\beta HV$ (where $\alpha=0$). Parameters below the dashed line are derived parameters.}
\def\arraystretch{1.7}
\begin{tabularx}{\textwidth}{Xcc} 
\hline\hline
Parameters                     & {\small TT,TE,EE+lowE + Pantheon} & {\small TT,TE,EE+lowE + Pantheon + BAO-RSD}         \\ \hline
$100\,\Omega_{b,0} h^2$         & $2.235\pm{0.015}$                 & $2.239\pm{0.014}$    \\
$100\,\theta_{*}$               & $1.04188\pm{0.00030}$  &  $1.04195\pm{0.00029}$  \\
$\tau$ & $0.0542_{-0.0081}^{+0.0076}$ & $0.0575_{-0.0086}^{+0.0073}$ \\ 
$\Omega_{c,0} h^2$   & $0.109_{-0.011}^{+0.013}$         & $0.1141_{-0.0087}^{+0.0094}$ \\
$\mathrm{ln}(10^{10}A_s)$       & $3.044_{-0.017}^{+0.016}$         & $3.050_{-0.018}^{+0.016}$     \\
$n_{s}$                         & $0.9643\pm{0.0045}$               & $0.9658\pm{0.0040}$    \\
$\beta$                         & $0.10\pm{0.11}$                   & $0.051_{-0.087}^{+0.089}$  \\
\hdashline
$\Omega_{V,0}$                  & $0.715_{-0.035}^{+0.034}$         & $0.703_{-0.025}^{+0.026}$   \\
$H_{0}$                         & $68.3\pm{1.1}$                    & $68.05_{-0.75}^{+0.76}$      \\
$\sigma_{8}$                    & $0.883_{-0.099}^{+0.064}$         & $0.848_{-0.067}^{+0.050}$    \\
$S_{8}$                         & $0.852_{-0.038}^{+0.027}$         & $0.838_{-0.028}^{+0.023}$     \\
\hline 
\end{tabularx}

\end{table}

\begin{figure}
\centering
\includegraphics[width=\textwidth]{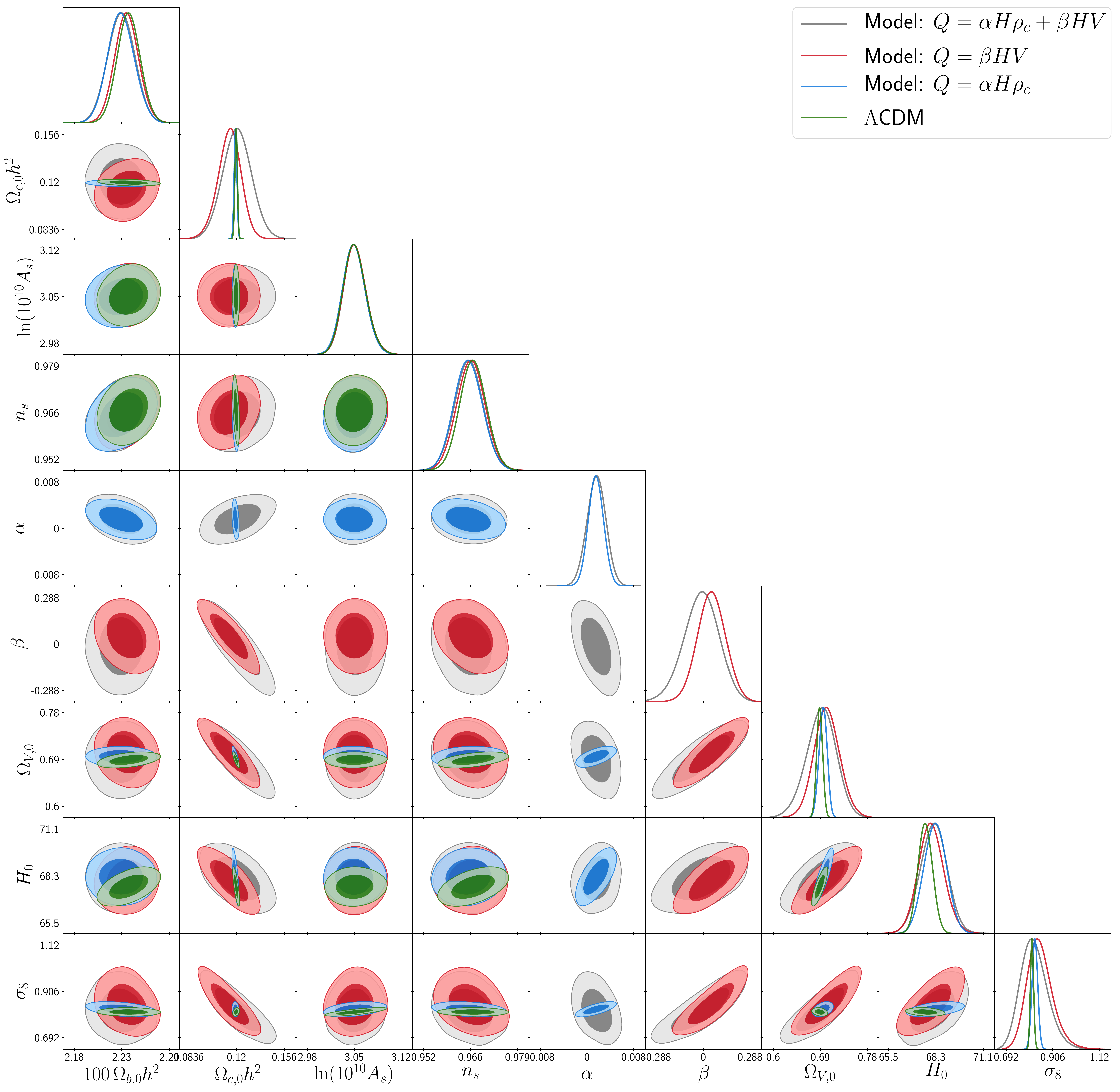}
\caption{\label{fig:fulltriangle-alpha+beta}1D and 2D marginal distributions for parameters in the three interaction models $Q=\alpha H\rho_c$ (in blue), $Q=\beta HV$ (in red), and $Q=\alpha H\rho_c+\beta HV$ (in grey) compared against the $\Lambda$CDM model parameter constraint (in green). In each case the marginal distributions are shown using the Planck 2018 CMB data (TT, TE and EE and lowE), plus Pantheon and BAO-RSD data. }
\end{figure}

Finally in Figure~\ref{fig:fulltriangle-alpha+beta} we show the marginal distribution for model parameters for all three of our interaction models, including the general linear interaction model $Q=\alpha H\rho_c+\beta HV$, using all three observational datasets from Planck 2018, Pantheon and BAO-RSD.
The resulting bounds on model parameters are given in Table~\ref{tab:allmodels}.
As one might expect there is some degeneracy between the parameters $\alpha$ and $\beta$; the early-time growth of the vacuum energy due to $\alpha>0$ can partly be compensated by late-time vacuum decay due to $\beta<0$, and vice versa. However the second and higher CMB acoustic peaks are sensitive to the decay or growth of dark matter ($\alpha\neq0$) around the time of recombination, in addition to the low-redshift constraints. Hence the allowed range for $\alpha$ is only slightly increased, $\alpha=0.0016^{+0.0017}_{-0.0016}$, when we allow for $\beta\neq0$. On the other hand the allowed range for $\beta$, which is only constrained by the low-redshift datasets, increases and shifts by about half of a standard deviation towards negative values, $\beta=-0.01^{+0.12}_{-0.11}$.

\begin{table}[h]
\centering
\caption{\label{tab:allmodels}Parameter constraints using Planck 2018 TT,TE,EE+lowE plus Pantheon and BAO-RSD datasets. Parameters below the dashed line are derived parameters.}
\def\arraystretch{1.7}
\begin{tabularx}{\textwidth}{Xcccc}
\hline\hline
Parameter                                & $\Lambda$CDM                 & $Q=\alpha H\rho_c$           & $Q=\beta HV$                 & {\small$Q=\alpha H\rho_c + \beta HV$} \\ \hline
$100\,\Omega_{b,0} h^2$\dotfill & $2.242\pm{0.014}$            & $2.233\pm{0.016}$            & $2.239\pm{0.014}$            & $2.233_{-0.016}^{+0.015}$        \\
$100\,\theta_{*}$\dotfill & \small$1.04197\pm{0.00029}$ & \small$1.04191\pm{0.00029}$ & \small$1.04195\pm{0.00029}$ & \small$1.04192_{-0.00029}^{+0.00030}$ \\
$\tau$\dotfill & $0.05775_{-0.0084}^{+0.0074}$ & $0.0559_{-0.0083}^{+0.0075}$ & $0.0575_{-0.0086}^{+0.0073}$ & $0.05631_{-0.0083}^{+0.0074}$ \\
$\Omega_{c,0} h^2$\dotfill    & $0.1194\pm{0.0010}$            & $0.1189\pm{0.0010}$            & $0.1141_{-0.0087}^{+0.0094}$ & $0.120\pm{0.011}$               \\
$\mathrm{ln}(10^{10}A_s)$\dotfill        & $3.050_{-0.017}^{+0.016}$              & $3.049_{-0.017}^{+0.016}$    & $3.050_{-0.018}^{+0.016}$       & $3.050_{-0.017}^{+0.015}$         \\
$n_{s}$\dotfill                          & $0.9665_{-0.0039}^{+0.0038}$ & $0.9651_{-0.0041}^{+0.0039}$ & $0.9658\pm{0.0040}$            & $0.9652_{-0.0042}^{+0.0041}$     \\
$\alpha$\dotfill                         & -                            & $0.0016\pm{0.0014}$          & -                            & $0.0016_{-0.0016}^{+0.0017}$   \\
$\beta$\dotfill                          & -                            & -                            & $0.051_{-0.087}^{+0.089}$    & $-0.01_{-0.11}^{+0.12}$       \\
\hdashline
$\Omega_{V,0}$\dotfill          & $0.6890_{-0.0058}^{+0.0059}$            & $0.6954_{-0.0076}^{+0.0080}$ & $0.703_{-0.025}^{+0.026}$    & $0.691_{-0.027}^{+0.030}$                \\
$H_{0}$\dotfill                          & $67.69_{-0.44}^{+0.43}$      & $68.27\pm{0.65}$             & $68.05_{-0.75}^{+0.76}$      & $68.17\pm{0.79}$                 \\
$\sigma_{8}$\dotfill                     & $0.811_{-0.008}^{+0.007}$    & $0.824_{-0.014}^{+0.013}$    & $0.848_{-0.067}^{+0.050}$    & $0.821_{-0.070}^{+0.053}$         \\
$S_{8}$                                  & $0.824_{-0.013}^{+0.012}$    & $0.828\pm{0.013}$            & $0.838_{-0.028}^{+0.023}$    & $0.827_{-0.030}^{+0.024}$     \\
\hline                                                  
\end{tabularx}

\end{table}



%
%
%


%


\section{Discussion}

\label{sec:discussion}
The marginal distributions in Figure~\ref{fig:fulltriangle-alpha+beta} illustrate how some cosmological parameters, which are primarily constrained by the CMB data, such as the baryon density, $\Omega_{b,0}h^2$, or the amplitude of the primordial scalar power spectrum, $A_s^2$, remain just as tightly constrained in the presence of the interactions. However, bounds on other parameters, such as the dark matter density $\Omega_{c,0}h^2$, may be substantially relaxed when one allows for interactions due to the previously noted degeneracies in the CMB spectra. We have seen that the interaction parameter $\beta$ is unconstrained by CMB data alone and, as a result, parameters which are degenerate with $\beta$ in the CMB data ($H_0$ and $\Omega_{c,0}h^2$) are only constrained when we include low-redshift datasets (Pantheon and BAO-RSD). The resulting constraints are then much weaker. For $Q=\beta HV$, for example, we find $\Omega_{c,0}h^2=0.114\pm0.009$ compared with $\Omega_{c,0}h^2=0.119\pm0.001$ for $\Lambda$CDM. In the general linear model, $Q=\alpha H\rho_c+\beta HV$, the uncertainty increases slightly to $\Omega_{c,0}h^2=0.114\pm0.011$.
While this broadens the allowed range for some parameters, none of these parameters differ by more than one standard deviation from their $\Lambda$CDM values when we allow for non-zero interactions.

According to the latest Hubble constant measurement by Hubble Space Telescope and the SH0ES Team~\cite{Riess:2021jrx}, the baseline result of the Hubble constant in our local universe is $H_0=73.04\pm1.04\;\mathrm{km\,s^{-1}Mpc^{-1}}$. This makes the discrepancy with respect to the value of $H_0$ inferred solely from the Planck 2018 CMB data within the $\Lambda$CDM model, $H_0=67.32\pm0.54\;\mathrm{km\,s^{-1}Mpc^{-1}}$ (see Table~\ref{tab:P18_six_param}), very nearly $5\sigma$. The inclusion of Pantheon and BAO-RSD datasets only tightens the bound slightly in the context of the $\Lambda$CDM cosmology to give $H_0=67.69^{+0.43}_{-0.44}\;\mathrm{km\,s^{-1}Mpc^{-1}}$ (see Table~\ref{tab:allmodels}).
Other independent distant-ladder estimates of $H_0$ also give higher values, for example based on the Tip of Red Giant Branch method~\cite{2019ApJ...882...34F} gives $H_0=69.8\pm0.8\pm1.7\;\mathrm{km\,s^{-1}Mpc^{-1}}$ (including an estimate of the systematic errors).

We have seen that in the presence of a dark matter-vacuum interaction there is a degeneracy between the interaction strength and other model parameters, including the value of the Hubble constant, when we consider only CMB data. This alleviates the tension between local measurements of the Hubble constant, such as the SH0ES measurement, and the value inferred from CMB data. For the $Q=\alpha H\rho_c$ interaction model (with $\beta=0$), where the second and higher acoustic peaks is sufficient to break the approximate degeneracy, the much weaker CMB bounds on $H_0$ reduce the tension to 2.3$\sigma$, even though the CMB data in this model actually have a mild preference for negative $\alpha$ and a lower value for $H_0$. For $\beta\neq0$ the interaction strength is not constrained by CMB data alone and thus the possibility of large values of $H_0$ compatible with the SH0ES result with $\beta>0$ is not suppressed.

The inclusion of Pantheon and/or BAO-RSD data, at lower redshifts than the CMB, breaks the degeneracy between the interaction strength and $H_0$. By constraining the interaction strength, the allowed range of $H_0$ is also bounded. The uncertainty on $H_0$ inferred from the combination of CMB, Pantheon and BAO-RSD datasets is increased in the presence of an interaction, reducing the $H_0$-tension slightly, but it remains close to 4$\sigma$
($3.9\sigma$ for both $\alpha\neq0$ and $\beta\neq0$ models, and $3.7\sigma$ for the general linear model with $\alpha\neq0$ and $\beta\neq0$).  

It is also interesting to consider the apparent tension between the amplitude of density fluctuations, $\sigma_8$, inferred from measurements by the Planck CMB experiment in $\Lambda$CDM cosmology and low-redshift observations. Weak lensing surveys are sensitive to the combination
$S_8\equiv\sigma_8\sqrt{\Omega_{m,0}/0.3}$. The recent $S_8$ measurement by KiDS-1000 survey reported $S_8=0.766_{-0.014}^{+0.020}$~\cite{Heymans:2020gsg}, while the value reported by the Dark Energy Survey is $S_8=0.776\pm{0.017}$~\cite{DES:2021wwk}. Both are well below the valued inferred from Planck in $\Lambda$CDM by about $2.7\sigma$ and $2.3\sigma$ respectively. 

\begin{figure}
\centering
\includegraphics[width=0.8\textwidth]{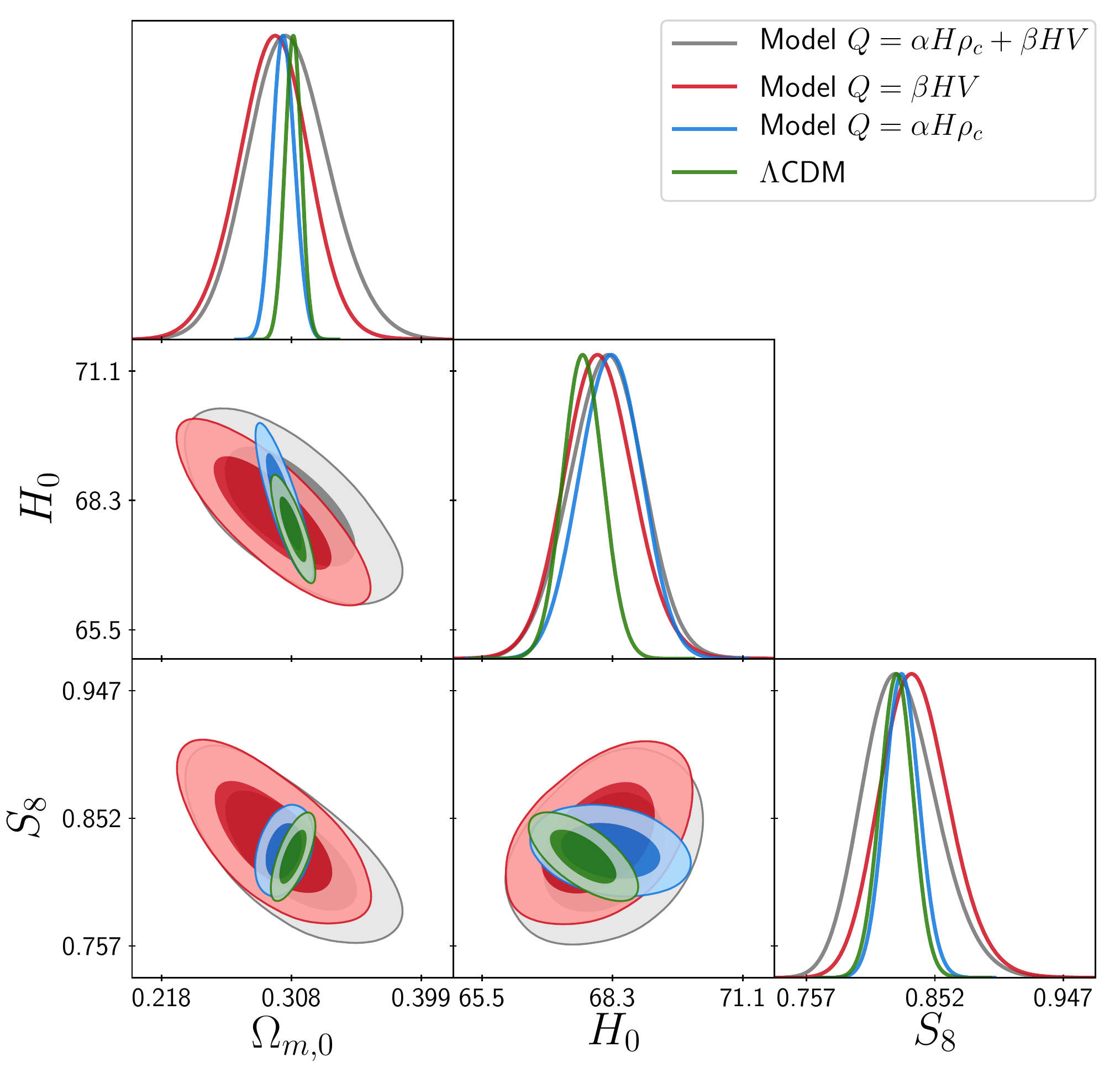}
\caption{\label{fig:S8}1D and 2D marginalised 
distributions for  
$S_8$,
$\Omega_{m,0}$ and $H_0$ for interacting vacuum models compared against $\Lambda$CDM (in green) using Planck 2018 TT, TE, EE and lowE, Pantheon and BAO-RSD data.}
\end{figure}

In Figure~\ref{fig:S8} we show the 1D and 2D marginalised distributions for $S_8$, $\Omega_{m,0}$ and $H_0$ where we use Pantheon and BAO-RSD data in addition to Planck 2018 data to constrain the interaction strength. 
In an interaction model with $Q=\alpha H\rho_c$ (and $\beta=0$) the matter density and hence $S_8$ remains tightly constrained. The data slightly prefer a positive value for $\alpha$ which corresponds to higher values of $H_0$ and slightly higher values for $S_8$ (slightly increasing the tension with KiDS-1000 to $2.9\sigma$).
On the other hand for $Q=\beta HV$ (and $\alpha=0$) the matter density and hence $S_8$ are less tightly constrained. Positive values for $\beta$ allow for higher $H_0$, but again this corresponds to higher values of $S_8$ (though the tension with KiDS-1000 is reduced somewhat to $2.4\sigma$).
In both cases it seems impossible to reconcile higher $H_0$ with smaller $S_8$. This is typical of many attempts to resolve these tensions found in $\Lambda$CDM \cite{Anchordoqui:2021gji}.

The most promising case may be the general linear interaction where $Q=\alpha H\rho_c+\beta HV$. In this case the presence of $\beta\neq0$ weakens the bounds on both $H_0$ and $S_8$ but simultaneously allowing $\alpha\neq0$ extends the allowed range of $S_8$ to lower values. 
The combination of $\alpha>0$ and $\beta<0$ allows both higher $H_0$ and lower $S_8$, reducing the tension with KiDS-1000 to $1.9\sigma$.


\section{Conclusion}
\label{conclusions}

In this paper we have explored observational constraints on cosmological models with an interacting vacuum, $\dot{V}=Q$ with a linear interaction, $Q=\alpha H\rho_c + \beta HV$, in which the energy exchange per Hubble time is proportional to the dark matter and/or vacuum energy density. 

We used linear perturbation theory to analyse the evolution of CMB anisotropies and density inhomogeneities. 
We have considered a geodesic model for the dark matter where the energy-momentum transfer, to or from the vacuum, follows the dark matter 4-velocity. Physically, this means that the dark matter still clusters, freely falling under gravity. This ensures that there is a synchronous frame comoving with the dark matter, in which the vacuum is spatially homogeneous, although time dependent. This simplifies the analysis of the CMB and large-scale structure, and we review the key equations for the growth of the linear density perturbations in Section~\ref{sec:cpt}. 
We modified the Einstein-Boltzmann code, \texttt{CLASS}, and used observational data including CMB temperature and polarisation and their cross correlation spectra, Type Ia Supernova data, Baryon Acoustic Oscillations and redshift-space distortions, to perform a Monte Carlo Markov Chain analysis to explore parameter constraints using a Metropolis-Hasting algorithm by using the \texttt{MontePython} code.

The presence of a non-zero energy transfer between dark matter and the vacuum leads to strong degeneracies in CMB power spectra, substantially relaxing bounds on some model parameters, such as $H_0$, with respect to the non-interacting ($\Lambda$CDM) case. The CMB data primarily constrains the angular scale of sound horizon at recombination, $\theta_*$, through the position of the first acoustic peak, and the matter density at recombination, $\omega_{c*}$, through the heights of the acoustic peaks. The remaining degeneracies seen in sub-section~\ref{ssec:results} closely match the analytic estimates given in Section~\ref{sec:cmb_aniso}. These degeneracies can be broken by including low-redshift data including type-Ia supernovae and BAO+RSD data.
In Section~\ref{sec:rsd} we emphasise the care that needs to be taken when interpreting redshift-space distortions in interacting models. The presence of an interaction means that the growth rate of structure, $f$ in Eq.~\eqref{deff}, differs from $f_\mathrm{rsd}$ in Eq.~\eqref{deffrsd} that is inferred from peculiar velocities and hence redshift-space distortion. 

We separately considered interaction models where $\alpha\neq0$, $\beta\neq0$ and the general case where $\alpha$ and $\beta$ are both non-zero.
Observational constraints in a growing (or decaying) vacuum energy model, $Q=\beta HV$ with $\beta>0$ (or $\beta<0$) have been previously studied a number of times ~\cite{Salvatelli:2014zta,Martinelli:2019dau}. Since the vacuum only becomes significant with respect to the dark matter density at late times, the dark matter can be treated as effectively non-interacting at high redshift. The physics of recombination is unchanged leading to a degeneracy between the interaction strength, $\beta$, and the present dark matter density, $\omega_c$, and Hubble constant, $H_0$, so long as the acoustic angular scale remains fixed. We can see in the right-hand panel of Fig.~\ref{fig:Cls_beta} that the CMB spectra are almost unchanged after fixing the angular acoustic scale at recombination, $\theta_{*}$. We break this degeneracy by including measurements of the late-time expansion history, using type-Ia supernovae and BAO+RSD data, to find $\beta=0.051^{+0.089}_{-0.087}$ similar to previous bounds on $|\beta|\lesssim 0.1$~\cite{Salvatelli:2014zta,Martinelli:2019dau}.
While a growing vacuum energy, $\beta>0$, still allows for larger values of $H_0$, hence reducing the tension between CMB and distance-ladder estimates of $H_0$, we observe that larger values of $H_0$ and associated with larger values of $S_8$, increasing the tension with weak-lensing data.

A decaying (or growing) dark matter model, $Q=\alpha H\rho_c$ with $\alpha>0$ (or $\alpha<0$), modifies the evolution of the Universe at both early and late times. There is again a degeneracy between the interaction strength and the present dark matter density such that the angular acoustic scale remains fixed. However, since any non-zero interaction affects the evolution of the dark matter density around the time of recombination, the impact of this on the higher peaks in the angular power spectrum, illustrated in Figure~\ref{fig:Cls_alpha}, is sufficient to break the degeneracy using CMB data alone. Planck 2018 data show a mild preference for negative values of $\alpha=-0.0077^{+0.0069}_{-0.0071}$, corresponding to lower values of $H_0$ compared with $\Lambda$CDM constraints, but with a large uncertainty, $H_0=62.2^{+4.2}_{-4.9}$.
The uncertainties in both $\alpha$ and $H_0$ are both more than halved by the inclusion of Pantheon (type-Ia supernovae) data at low-redshift, and halved again by including BAO+RSD data from eBOSS. The RSD data in particular introduce a preference for positive $\alpha=0.0016\pm0.0014$
and higher values for $H_0=68.27\pm0.65$. Recently, Sol\'{a} Peracaula et al~\cite{SolaPeracaula:2021gxi} found a similar uncertainty using CMB, supernovae and large-scale structure data for a closely related model with $Q=-3\nu_{\rm eff}H\rho_m$, but with a slightly different mean value given by $\nu_{\rm eff}=0.0002\pm0.0004$.

The eBOSS measurement of $f_{\rm rsd}\sigma_8$ at $z_{\rm eff}=1.48$ from quasar data plays an important role in our constraints on $\alpha$. This favours higher values for $f_{\rm rsd}\sigma_8$ at this redshift compared with the standard $\Lambda$CDM prediction. In the $Q=\beta HV$ model, even large variations in $\beta$ have little effect on $f_{\rm rsd}\sigma_8$ at $z\gtrsim1$ once we fix the acoustic angular scale and dark matter density at recombination (see Figure~\ref{fig:fsigma8_beta}). $\beta$ only changes the predictions for $f_{\rm rsd}\sigma_8$ at low redshifts. However in the $Q=\alpha H\rho_c$ model one obtains significantly higher values for $f_{\rm rsd}\sigma_8$ at $z\gtrsim1$ for $\alpha>0$, as seen in Figure~\ref{fig:fsigma8_alpha}. The eBOSS quasar data point shifts the preferred parameter values towards positive $\alpha$ and higher $H_0$, without significantly affecting the prediction for $S_8$, as shown in Figure~\ref{fig:S8}.

Interacting vacuum cosmologies offer simple predictive models in which to study the $H_0$ and $S_8$ tensions present when attempting to reconcile CMB constraints with low-redshift probes. To be able to simultaneously increase $H_0$ and reduce $S_8$ with respect to the $\Lambda$CDM values inferred from CMB constraints, we find that we need to consider both $\alpha$ and $\beta$ to be non-zero, as seen in Figure~\ref{fig:S8}.
Previous studies have argued that the scenarios where an interaction switches on only below some threshold redshift may be preferred by the data~\cite{Salvatelli:2014zta,Martinelli:2019dau,SolaPeracaula:2021gxi}. This may be mimicked in our general linear interaction model with $\alpha\neq0$ and $\beta\neq0$, where the different time dependence of the dark matter and vacuum energy densities lead to a more complex redshift-dependent interaction. 
%

We have stressed the important role played by observations of redshift-space distortions and the need to consistently interpret these observations when constraining model parameters if the growth rate of density perturbations is driven by interactions with the vacuum as well as the divergence of the velocity field. To fully exploit the potential of large-scale structure to constrain such interactions we will need go beyond the linear regime, which remains an challenge for future studies of interacting dark matter and dark energy.

%
%

{\em Note added:} While this paper was in the final stages of preparation Ref.~\cite{Yang:2022csz} appeared which considers interacting dark energy models including an interacting vacuum scenario (which they refer to as $\xi_0{\bf IVS}A$) that coincides with our model where $Q=\beta HV$. Their analysis using the same CMB plus Pantheon data matches our own constraints very closely in this case. Their results including BAO appear to be similar to ours, but they do not include RSD data, so the bounds cannot be directly compared.

\section*{Acknowledgements}

The authors are grateful to Thomas Tram for advice on the Einstein-Boltzmann code, \texttt{CLASS}. 
We thank Eva-Maria Mueller and Antony Lewis for useful discussions about BAO-RSD data.
CK is funded by the Royal Thai Government. 
HA, JS and DW are supported by the Science and Technology Facilities Council grants ST/S000550/1 and ST/W001225/1.
Supporting research data are available on reasonable request from the corresponding author.
For the purpose of open access, the authors have applied a CC BY public copyright licence to any Author Accepted Manuscript version arising.

\bibliographystyle{JHEP}
\bibliography{ref21}

\end{document}